\documentclass[12pt]{article}

\usepackage{latexsym}
\usepackage[utf8]{inputenc}
\usepackage[T1]{fontenc}
\usepackage[normalem]{ulem} 
\usepackage{epsfig,amssymb,euscript,slashed}
\usepackage{amsmath}
\usepackage{array,calc,epsfig}
\usepackage{braket}
\usepackage{xcolor}
\usepackage{float}
\usepackage{bm}
\usepackage{bbm}
\usepackage{fancybox}
\usepackage{comment}

\usepackage{hyperref}

\numberwithin{equation}{section}

\usepackage{cite}

\makeatletter
\def\blfootnote{\gdef\@thefnmark{}\@footnotetext}
\makeatother

\def\Re           {{\rm Re\hskip0.1em}}

\begin{document}

\title{
	\vspace{-3.0cm}\begin{flushright}{\scriptsize DFPD-2018/TH/02}\end{flushright}
	\hfill\\
	\hfill\\
On the dynamical origin of parameters in $\mathcal{N}=2$ Supersymmetry}
\author{Niccol\`o Cribiori$^a$ and Stefano Lanza$^{b}$}

\date{}

\maketitle

\vspace{-1.5cm}

\begin{center}
	\vspace{1.0cm}
	\textit{$^a$ Institute for Theoretical Physics, TU Wien,\\ Wiedner Hauptstrasse 8-10/136, A-1040 Vienna, Austria
		\\
		\vspace{0.5em}
		$^b$ Dipartimento di Fisica e Astronomia ``Galileo Galilei'',  Universit\`a degli Studi di Padova\\
		\& I.N.F.N. Sezione di Padova, Via F. Marzolo 8, 35131 Padova, Italy}
	\blfootnote{e-mails: \href{mailto:niccolo.cribiori@tuwien.ac.at}{niccolo.cribiori@tuwien.ac.at}, \href{mailto:stefano.lanza@pd.infn.it}{stefano.lanza@pd.infn.it} }
\end{center}

\vspace{1.5cm}

\begin{abstract}
We formulate $\mathcal{N}=2$ global supersymmetric Lagrangians of self-interacting vector multiplets in terms of variant multiplets, whose non-propagating fields are replaced with gauge three-forms. Setting the three-forms on-shell results in a dynamical generation of the parameters entering the scalar potential. As an application, we study how gauge three-forms may determine the partial breaking of $\mathcal{N}=2$ supersymmetry and how they affect the low energy effective description.
\end{abstract}

\thispagestyle{empty}

\newpage

\tableofcontents

\newpage


\section{Introduction and motivation}

In four spacetime dimensions gauge three-forms do not carry any propagating degrees of freedom, nevertheless they can induce non-trivial physical effects. Their importance was recognized in cosmology, where they have been employed to provide a dynamical way to generate the cosmological constant  \cite{Hawking:1984hk,Brown:1987dd,Brown:1988kg,Duncan:1989ug,Bousso:2000xa}. More recently, gauge three-forms have been used as tools to investigate gauge axionic shift symmetries, both for introducing new inflationary models  \cite{Kaloper:2008fb,Kaloper:2011jz,Marchesano:2014mla,Dudas:2014pva,Valenzuela:2016yny} and for addressing the strong $CP$-problem \cite{Dvali:2005an,Dvali:2004tma,Dvali:2005zk,Dvali:2013cpa,Dvali:2016uhn,Dvali:2016eay}. Indeed gauge three-forms have been embedded in theories enjoying global or local supersymmetry \cite{Duff:1980qv, Stelle:1978ye,Aurilia:1980xj,Gates:1980ay,Gates:1980az,Buchbinder:1988tj,Binetruy:1996xw,Ovrut:1997ur,Kuzenko:2005wh,Nishino:2009zz,Duff:2010vy,Groh:2012tf,Aoki:2016rfz,Nitta:2018yzb}. In this context, they might provide new insights for nonlinear realizations of supersymmetry \cite{Farakos:2016hly,Buchbinder:2017vnb,Buchbinder:2017qls,Kuzenko:2017vil} and for their natural coupling to membranes which, firstly explored in cosmology \cite{Brown:1987dd,Brown:1988kg}, was also extended in supersymmetric theories \cite{Ovrut:1997ur,Bandos:2010yy,Bandos:2011fw,Bandos:2012gz,Bandos:2018gjp}.

More generically, gauge three-forms can be seen as counterparts of constant parameters appearing in four-dimensional effective theories, for which they provide a dynamical origin. For example, in \cite{Bousso:2000xa,Bielleman:2015ina,Farakos:2017jme,Farakos:2017ocw,Carta:2016ynn,Herraez:2018vae} the parameters appearing in the $F$-term potential of Type II string theory compactified over Calabi-Yau three-folds were interpreted as expectation values of the field-strengths of some gauge three-forms. 
It is in fact of prominent importance that all the parameters of the effective field theories stemming from string theory, but the string length, can be understood as expectation values of some fields and, in this respect, the presence of gauge three-forms may come to help.

In this spirit, in \cite{Farakos:2017jme,Farakos:2017ocw} four-dimensional theories with $\mathcal{N}=1$ global and local supersymmetry have been analysed. A procedure has been given to construct Lagrangians encoding gauge three-forms, which are on-shell equivalent to a generic class of chiral models. In particular, the chiral superfields in the latter are substituted in the former by variant versions, which contain gauge three-forms as highest components. Once such three-forms are integrated out, parameters are introduced which contribute to the superpotential of the given chiral model.

In this work we elaborate on these results in two directions. First of all, we extend the procedure of \cite{Farakos:2017jme,Farakos:2017ocw} in order to incorporate also $\mathcal{N}=1$ vector superfields. This will allow us to construct models in which the Fayet--Iliopoulos parameter is generated dynamically. Secondly, we investigate $\mathcal{N}=2$ rigid supersymmetric Lagrangians for $\mathcal{N}=2$ vector multiplets and we reformulate them in terms of new variant multiplets containing gauge three-forms as non-propagating degrees of freedom. Since an $\mathcal{N}=2$ vector superfield contains three real auxiliary fields, as a consequence of the procedure we propose, for any such a superfield three real parameters are going to be generated dynamically. On the contrary to the $\mathcal{N}=1$ case, in which only specific parameters of a given superpotential can be interpreted as vacuum expectation values of the field strength of gauge three-forms, the approach in the $\mathcal{N}=2$ case is covering a more general situation, in which the entire potential has a dynamical origin.
In other words, an $\mathcal{N}=2$ off-shell Lagrangian for variant vector multiplets will contain no parameters at all, since they are all going to be introduced when integrating out the non-propagating gauge three-forms.

As an application, the partial breaking of global supersymmetry is reviewed \cite{Antoniadis:1995vb, Ferrara:1995xi,Bagger:1996wp,Fujiwara:2004kc,Fujiwara:2005hj,Ambrosetti:2009za,Kuzenko:2015rfx,Antoniadis:2017jsk,Antoniadis:2018blk,Farakos:2018aml} and reconstructed in the formulation with variant multiplets. The mutual orientation of the gauge three-forms will dictate whether the original, off-shell $\mathcal{N}=2$ supersymmetry may be partially broken to $\mathcal{N}=1$, once the gauge three-forms are set on-shell. We then construct an effective field theory with partially broken supersymmetry, which leads to an action of the Born--Infeld type \cite{Bagger:1996wp,Rocek:1997hi}, in which the supersymmetry breaking parameter appearing in front of the Lagrangian is generated dynamically. 
The information contained in this effective description is entirely encoded into boundary terms, which are necessary in the presence of gauge three-forms.

Throughout this work we use the superspace conventions of \cite{Wess:1992cp} and, even in the case of extended supersymmetry, most of the calculation are performed at the $\mathcal{N}=1$ superspace level for convenience.


\section{Three-forms in $\mathcal{N}=1$ global supersymmetry}
\label{sec:N1case}

In four dimensions, gauge three-forms can be accommodated inside the auxiliary components of $\mathcal{N}=1$ superfields. Variant formulations of chiral and vector superfields have been constructed in \cite{Gates:1983nr}, where the usual complex or real scalar auxiliary fields are exchanged with (the Hodge-dual of) field strengths of these three-forms. In this section, following the method introduced in \cite{Farakos:2017jme}, Lagrangians are constructed for variant superfields, which are on-shell equivalent to the standard ones. One of the advantages of dealing with these alternative Lagrangians resides in the fact that the parameters appearing inside them, as for example in the superpotential or in the Fayet--Iliopoulos term, are going to be generated dynamically as vacuum expectation values.
In addition, the variant $\mathcal{N}=1$ superfields introduced in this section will be essential ingredients to construct variant $\mathcal{N}=2$ superfields in the next sections.

\subsection{Double three-form chiral multiplets and dynamical generation of the linear superpotential}
\label{sec:N1caseC}

Three-form multiplets are chiral multiplets whose non-propagating degrees of freedom are encoded into gauge three-forms, rather than complex scalar fields. Both single and double three-form multiplets can be constructed \cite{Gates:1980ay,Gates:1980az,Gates:1983nr} in which, respectively, one or two real non-propagating degrees of freedom are replaced by three-forms. In \cite{Farakos:2017jme} it was shown how to dynamically pass from ordinary chiral multiplets to three-form multiplets at the Lagrangian level. The construction is reviewed here for the case of the double three-form multiplet and by looking at a simple example.

Consider a chiral multiplet $X$. It can be expanded in chiral coordinates in superspace as 
\begin{equation}
X = \varphi + \sqrt 2 \, \theta \psi +\theta^2 f,
\end{equation}
where $\varphi$ is a complex scalar, $\psi$ a Weyl fermion and $f$ a complex scalar auxiliary field.\footnote{The reader is referred to appendix \ref{app:Super} for more details on the component structure of the superfields introduced here and in the following.} The most general Lagrangian, up to two derivatives, which can be constructed solely in terms of this ingredient is
\begin{equation}
\label{N1L}
\mathcal{L}= \int d^4\theta \,K(X, \bar X) + \left(\int d^2 \theta\, W(X) + c.c\right)\,,
\end{equation}
where $K(X,\bar X)$ is the K\"ahler potential and $W(X)$ is the superpotential, which is a holomorphic function of $X$.
Without loss of generality, the superpotential can be rewritten as  
\begin{equation}
W(X) = c\, X + \hat W(X)\,,
\end{equation}
where $c$ is a complex constant and the function $\hat W(X)$ is holomorphic.
The bosonic components of \eqref{N1L} are
\begin{equation}
\label{N1bosA}
\mathcal{L} \big|_{\text{bos}} = -K_{\varphi \bar \varphi} \partial_m \varphi \partial^m \bar \varphi + K_{\varphi \bar \varphi} f \bar f+ \left[(c + \hat W_\varphi(\varphi)) f + \text{c.c.}\right]\,
\end{equation}
and, setting the auxiliary field $f$ on-shell
\begin{equation}
f = -\frac{\bar c+\bar{\hat{W}}_{\bar \varphi}}{K_{\varphi\bar\varphi}}\,,
\end{equation}
the Lagrangian becomes
\begin{equation}
\label{N1bosON}
\mathcal{L} \big|_{\text{bos, on-shell}} = -K_{\varphi \bar \varphi} \partial_m \varphi \partial^m \bar \varphi - \mathcal{V}(\varphi,\bar\varphi)\,,
\end{equation}
where the scalar potential is
\begin{equation}
\mathcal{V}(\varphi,\bar\varphi) = \frac{1}{K_{\varphi\bar\varphi}}\left|c + \hat W_\varphi(\varphi)\right|^2.
\end{equation}

As shown in \cite{Farakos:2017jme}, the Lagrangian \eqref{N1L} can be thought of as originating from a parent Lagrangian for the double three-form multiplet, in which the parameter $c$ is generated dynamically. To construct such a Lagrangian we can start from
\begin{equation}
\label{N1LB}
\mathcal{L}= \int d^4\theta K(X, \bar X) + \left(\int d^2 \theta\, \left(\Phi X + \frac14   \bar D^2 (\Sigma \bar \Phi) \right)+\int d^2 \theta\, \hat W(X) + {\rm c.c.}\right)\,,
\end{equation}
where, with respect to the \eqref{N1L}, the linear part of the superpotential has been promoted as
\begin{equation}
\int d^2\theta\, c\,X \rightarrow \int d^2 \theta\, \left( \Phi X + \frac14   \bar D^2 (\Sigma \bar \Phi)  \right)\,.
\end{equation}
Here $\Phi$ is a chiral superfield with no kinetic terms, which will ultimately play the role of Lagrange multiplier, while $\Sigma$ is a complex linear multiplet, namely a complex scalar multiplet which is constrained by
\begin{equation}
\label{ConS}
\bar D^2 \Sigma = 0.
\end{equation}
Its superspace expansion is
\begin{equation}
\begin{split}
\Sigma = &\sigma + \sqrt 2\theta \psi + \sqrt{2} \bar\theta \bar\rho - \theta \sigma_m \bar\theta \mathcal{B}^{m} + \theta^2 \bar s+ \sqrt 2 \theta^2\bar\theta \bar\zeta \\
&-\frac{i}{\sqrt{2}} \bar\theta^2 \theta \sigma^m \partial_m \bar\rho + \theta^2\bar\theta^2 \left(\frac{i}{2} \partial_m \mathcal{B}^{m} -\frac14 \Box \sigma \right),
\end{split}
\label{CompS}
\end{equation}
where $\sigma$ and $s$ are complex scalar fields,  $\psi$, $ \rho$ and $\zeta$ Weyl fermions, while the complex vector $\mathcal{B}^m$ can be interpreted as being the Hodge-dual of a complex three-form $\mathcal{B}_3 = \frac{1}{3!}\mathcal{B}_{mnl} dx^m \wedge dx^n \wedge dx^l$ as
\begin{equation}
\mathcal{B}^m=\frac{1}{3!}\epsilon^{mnlp}\mathcal{B}_{nlp}\,.
\end{equation}

As a consistency check, it is possible to integrate out $\Sigma$ from \eqref{N1LB} and recover the original Lagrangian \eqref{N1L}. Since the superfield $\Sigma$ is constrained, it is not possible to take directly its variation. 
However, the constraint \eqref{ConS} can be solved as
\begin{equation}
\Sigma = \bar D_{\dot \alpha} \bar \Psi^{\dot \alpha}\,,
\end{equation}
with $\bar \Psi^{\dot \alpha}$ an unconstrained spinorial superfield. The variation with respect to $\bar \Psi^{\dot \alpha}$ produces
\begin{equation}
\bar D_{\dot \alpha} \bar \Phi = 0
\end{equation}
and, since $\Phi$ chiral, the only possibility is that
\begin{equation}
\label{N1solX}
\Phi=c\,,
\end{equation}
with $c$ an arbitrary complex constant. Plugging \eqref{N1solX} into \eqref{N1LB} we thus obtain the Lagrangian \eqref{N1L}.

On the other hand, it is possible to integrate out from \eqref{N1LB} both the Lagrange multiplier $\Phi$ and the chiral multiplet $X$. The variation with respect to $X$ gives the superspace equations of motion
\begin{equation}
\label{N1solXb}
\begin{split}
\Phi &= \frac14 \bar D^2 K_X - \hat W_X\,,
\end{split}
\end{equation}
while the variation with respect to the Lagrange multiplier replaces the old chiral superfield with a new one, which is expressed in terms of the complex linear multiplet $\Sigma$
\begin{equation}
\label{N1solPhi}
\begin{split}
X &= -\frac14 \bar D^2 \bar \Sigma \equiv S \,.
\end{split}
\end{equation}
The superfield $S$ is called \emph{double three-form multiplet} and it has been constructed dynamically from the Lagrangian \eqref{N1LB}.  It is chiral and it can be expanded in superspace as
\begin{equation}
S = \varphi^S + \sqrt 2 \, \theta \psi^S + \theta^2 f^S.
\end{equation}
Its components, in terms of those of $\Sigma$, are (see Table \ref{tab:comp1})
\begin{align}
\label{compSintermsofSigma1}
\varphi^S &= s,\\
\label{compSintermsofSigma2}
\psi^S_\alpha &= \zeta_\alpha + \frac{1}{2} i \sigma^m_{\alpha \dot \beta}\partial_m \bar \psi^{\dot \beta}\,, \\
\label{compSintermsofSigma3}
f^S &=-i{}^*\! \bar F_4 = -i \partial_m \bar{\mathcal{B}}^m,
\end{align}
with 
\begin{equation}
{}^*\!F_4 = \frac{1}{4!}\varepsilon^{klmn} F_{klmn},\qquad F_{klmn} = 4 \partial_{[k} \mathcal{B}_{lmn]}\,.
\end{equation}
In addition, with respect to the standard chiral superfield $X$, the multiplet $S$ is invariant under the shift 
\begin{equation}
\label{Sgauge}
\Sigma \rightarrow \Sigma + L_1 + i L_2,
\end{equation}
where $L_1$ and $L_2$ are real linear superfields. As a consequence, the complex three-form $\mathcal{B}_{klm}$ undergoes a gauge transformation of the type
\begin{equation}
\mathcal{B}_{klm} \rightarrow \mathcal{B}_{klm} + 3\partial_{[k} \left( \Lambda_1 + i \Lambda_2\right)_{lm]}\,.
\end{equation}
where $\Lambda_{1\,mn}$ and $\Lambda_{2\,mn}$ are components of real gauge two-forms. In this sense, the complex linear superfield $\Sigma$ contains a gauge three-form among its components. Moreover, with an appropriate gauge choice, it is possible to set $\psi_\alpha  = \rho_\alpha=0$ and the fermionic component of $S$ becomes $\psi^S_\alpha = \zeta_\alpha$. Therefore, the double three-form multiplet $S$ shares the same degrees of freedom as a chiral multiplet, even though its auxiliary component is not a complex scalar field, but it is the Hodge-dual of the field strength of the gauge three-form $\mathcal{B}_{klm}$.

Plugging \eqref{N1solXb} and \eqref{N1solPhi} into \eqref{N1LB}, the desired Lagrangian in terms of $S$ is obtained
\begin{equation}
\label{N1L3}
\mathcal{L}= \int d^4\theta\, K(S, \bar S) + \left(\int d^2 \theta\, \hat W(S) + c.c\right)+ \mathcal{L}_{\text{bd}}\,,
\end{equation}
where
\begin{equation}\label{N1Sbd}
\mathcal{L}_{\rm bd}=\frac14 \left(\int d^2\theta \bar D^2 - \int d^2\bar\theta D^2\right) \left[\left(\frac14  D^2 K_{\bar S} - \bar{\hat{W}}_{\bar S}\right) \Sigma\right]
+\text{c.c.} \, 
\end{equation}
are boundary terms which are necessary to ensure the correct variation of the action with respect to the gauge three-form \cite{Brown:1987dd, Brown:1988kg, Groh:2012tf}. The bosonic components of \eqref{N1L3} are
\begin{equation}
\label{N1bosB}
\mathcal{L} \big|_{\text{bos}} = -K_{S \bar S} \partial_m s \partial^m \bar s - \frac{1}{4!} K_{S \bar S} F^{klmn} \bar F_{klmn} + \left[-i \hat{W}_S(s) {}^*\!\bar F_4 + c.c.\right] + \mathcal{L}_{\text{bd}},
\end{equation}
where
\begin{equation}
\label{N1bosBbd}
\mathcal{L}_{\text{bd}} = \frac{1}{3!} \partial_k \left[i \mathcal{B}_{lmn} \left(-iK_{S \bar S} \bar F^{klmn} - \varepsilon^{klmn}\bar{\hat{W}}_S(s) \right)\right] + c.c.\,.
\end{equation}
As it can be shown from \eqref{N1bosB}, indeed the boundary terms \eqref{N1bosBbd} cancel those originating from the variation of the action defined by \eqref{N1bosB} with respect to the gauge three-form. 
The potential of the parent Lagrangian \eqref{N1bosB} can be obtained by setting the gauge three-form on-shell:
\begin{equation}
\label{N1bosFsola}
\partial_k \left[iK_{S \bar S} F^{klmn} - \varepsilon^{klmn}\hat{W}_S(s)\right] = 0
\end{equation}
whence
\begin{equation}
\label{N1bosFsol}
iF_{klmn} = \frac{c+\hat{W}_S(s)}{K_{S \bar S}} \varepsilon_{klmn}\,,
\end{equation}
with $c$ an arbitrary complex constant. Plugging the solution \eqref{N1bosFsol} in \eqref{N1bosB} we obtain exactly the model \eqref{N1bosON}. The advantage of starting from the Lagrangian \eqref{N1bosB}, rather than \eqref{N1bosA}, is that no supersymmetry breaking parameter appears. In fact, in \eqref{N1bosB} the complex constant $c$ is dynamically generated by solving the equation of motion for the gauge three-form. In other words, the constant $c$ has been promoted to the vacuum expectation value of the particular combination which appears in \eqref{N1bosFsola} or, equivalently, the choice of such a parameter has been traded for the specification of the boundary condition for the gauge three-form.

\begin{table}
\centering
\begin{tabular}{|c c c c|} 
\hline
Superfield & Spin-$0$ & Spin-$\frac12$ & Non-propagating\\ [0.5ex] 
\hline \hline
$X$ & $\varphi$ & $\psi_\alpha$ & $f$\\
$S$ & $s$ & $\psi^S_\alpha$ & $\mathcal{B}_{mnp}$\\
\hline
\end{tabular}
\caption{The off-shell degrees of freedom of the ordinary chiral superfield $X$ and of the double three-form multiplet $S$. The complex auxiliary field $f$ of $X$ is replaced by a complex gauge three-form in the variant version.}
\label{tab:comp1}
\end{table}

\subsection{Three-form vector multiplet and dynamical realization of the Fayet--Iliopoulos term}

\label{sec:N1caseV}

In this subsection, we extend the procedure of \cite{Farakos:2017jme} to the case of a Fayet--Iliopoulos parameter, which is going to be dynamically generated as vacuum expectation value of a real gauge three-form. The discussion is again performed at the Lagrangian level and a parent Lagrangian for a variant vector multiplet will be introduced. Such a multiplet, which has been previously constructed for example in \cite{Gates:1983nr, Antoniadis:2017jsk}, accomodates a real gauge three-form as non-propagating field, in analogy with \eqref{N1solPhi}.

Given a vector multiplet $V$, the minimal Lagrangian which can be built is
\begin{equation}
\label{VLagS}
\mathcal{L} =\left(\frac14 \int d^2\theta\, W^\alpha W_\alpha + \text{c.c.}\right)+ \xi \int d^4\theta\, V
\end{equation}
where $\xi \in \mathbb{R}$ is the so called Fayet--Iliopoulos parameter. Focusing only on the bosonic sector for simplicity, the components reads
\begin{equation}
\label{VLagOffSa}
\mathcal{L} |_{\text{bos}}= - \frac 14  F^{mn}F_{mn}+\frac12 {\rm D}^2  +\frac{\xi}{2}\, {\rm D}\,,
\end{equation}
where we have neglected the total derivative which involves the gauge field and defined $F_{mn} \equiv 2 \partial_{[m} v_{n]}$. Setting the auxiliary field ${\rm D}$ on-shell
\begin{equation}
{\rm D} = -\frac{\xi}{2}\,,
\end{equation}
we get
\begin{equation}
\label{VLagOnSa}
\mathcal{L}|_{\text{bos, on-shell}} =  - \frac 14  F^{mn}F_{mn}  -\frac{\xi^2}{8}
\end{equation}
with a constant, semi-positive definite potential $\mathcal{V} = \frac{\xi^2}{8}$.

In the spirit of the previous discussion, instead of considering \eqref{VLagS}, we can start from the parent Lagrangian
\begin{equation}
\begin{split}
\label{VLagMaster}
\mathcal{L} =&\frac14 \left(\int d^2\theta\, W^\alpha W_\alpha + \text{c.c.}\right) -\frac18 \int d^2\theta\,\bar D^2 (\Lambda V) -\frac18 \int d^2\bar\theta\, D^2 (\Lambda V) 
\\
&+\frac18 \left[ \int d^2\theta \bar D^2 (\Lambda \Sigma)+\int d^2\bar\theta D^2 (\Lambda \bar \Sigma)\right]\,,
\end{split}
\end{equation}
which is obtained by promoting the Fayet--Iliopoulos parameter $\xi$ to a real Lagrangian multiplier $\Lambda$ and conveniently adding new terms which contain the complex linear multiplet $\Sigma$ encoding the three-form.

First, we check that from this Lagrangian we get  the usual Lagrangian \eqref{VLagS}. This can be achieved by eliminating the dependence in \eqref{VLagMaster} on the gauge three-form. By integrating out the unconstrained spinorial superfields $\Psi^\alpha$ and $\bar\Psi^{\dot \alpha}$, such that $\Sigma = \bar D_{\dot \alpha} \bar \Psi^{\dot \alpha}$, we get
\begin{equation}
D_\alpha \Lambda = 0, \quad \bar D_{\dot\alpha} \Lambda = 0\,,
\end{equation}
whence $\Lambda$ is just a real constant $\xi$
\begin{equation}
\Lambda = \xi\,.
\end{equation}
Once inserted into \eqref{VLagMaster}, the ordinary vector multiplet Lagrangian \eqref{VLagS} is recovered. 

Let us now follow a second path in order to express the parent Lagrangian solely in terms of a new real three-form multiplet.
The variation of the Lagrangian \eqref{VLagMaster} with respect to the Lagrangian multiplier $\Lambda$ gives
\begin{equation}
\label{VsolV}
V = \frac{\Sigma + \bar\Sigma}{2}\equiv U,
\end{equation}
which is a real multiplet containing a gauge three-form as auxiliary degree of freedom \cite{Gates:1983nr, Antoniadis:2017jsk}. This can be understood as follows. We recall that $\Sigma$ contains in its expansion a complex vector $\mathcal{B}^m$
\begin{equation}
\frac14 \bar{\sigma}^{m\,\dot\alpha \alpha}
\left[D_\alpha,\bar{D}_{\dot{\alpha}}\right] \Sigma| = \mathcal{B}^m \equiv B^m + i C^m\,,
\end{equation}
where $B^m \equiv {\rm Re}\, \mathcal{B}^m$ and $C^m \equiv {\rm Im}\, \mathcal{B}^m$. As discussed before, we can interpret $C^m$ as the Hodge dual of a three-form $C_{mnp}$
\begin{equation}
C^m = \frac{1}{3!}\varepsilon^{mnpq} C_{npq}\,,
\end{equation}
whose four-form field strength $G_4$ components are defined as
\begin{equation}
G_{mnpq} = 4 \partial_{[m} C_{npq]}\,.
\end{equation}
Using \eqref{SigmaCom} we get
\begin{equation}
\label{Vvara}
\begin{split}
\frac14 \bar{\sigma}^{m\,\dot\alpha \alpha}\left[D_\alpha,\bar{D}_{\dot{\alpha}}\right] U| &=  B^m\\
\frac{1}{16}  D^2 \bar D^2 U| &= -\frac12 {}^*\! G_4 - \frac{i}{2} \partial_m B^m
\end{split}
\end{equation}
where we have introduced ${}^*\! G_4$, which is the Hodge-dual of the field strength $G_{mnpq}$ (see Table \ref{tab:comp2}).
Comparing \eqref{Vvara} with the usual projection of an ordinary real multiplet $V$
\begin{equation}
\label{Vvarb}
\begin{split}
\frac14 \bar{\sigma}^{m\,\dot\alpha \alpha}\left[D_\alpha,\bar{D}_{\dot{\alpha}}\right] V| &=  v^m\\
\frac{1}{16}  D^2 \bar D^2 V| &= \frac{\rm D}{2}-\frac{i}{2} \partial_mv^m
\end{split}
\end{equation}
we recognize that, in the variant formulation \eqref{Vvara}, the auxiliary field ${\rm D}$ of the ordinary vector multiplet is replaced with the Hodge-dual of the field strength of the three-form $C_{mnp}$, namely
\begin{equation}
{\rm D} \rightarrow  -\, {}^*\! G_4\,.
\end{equation}

Moreover, the Lagrangian \eqref{VLagMaster} is invariant under the shift
\begin{equation}
\label{Ugauge}
\Sigma \rightarrow \Sigma + \Phi + i L
\end{equation}
with $\Phi$ and $L$ being, respectively, an arbitrary chiral and real linear multiplet. This in turns induces a gauge transformation for the vector three-form multiplet of the standard form
\begin{equation}
U \to U + \Phi + \bar \Phi.
\end{equation}
and reflects on the gauge transformation for the three-form $C_{mnp}$ as
\begin{equation}
C_{mnp} \rightarrow C_{mnp} + 3\partial_{[m} \Lambda_{np]}
\end{equation}
where $\Lambda_{mn}$ is an arbitrary real gauge two-form. This enforces the interpretation of $C_{mnp}$ as components of a gauge three-form.

Therefore the multiplet\eqref{VsolV} is the counterpart of the chiral double three-form multiplet and, in the following, we will dub it \emph{real} (or \emph{vector}) \emph{three-form multiplet}.

\begin{table}
\centering
\begin{tabular}{|c c c c|} 
\hline
Superfield & Spin-$\frac12$ & Spin-$1$ & Non-propagating \\ [0.5ex] 
\hline \hline
$V$ & $\lambda$ & $v^m$ & ${\rm D}$\\
$U$ & $\lambda_U$ & $B^m$ & $C_{mnp}$\\
\hline
\end{tabular}
\caption{The fundamental off-shell component fields of the ordinary real superfield $V$ and of the real three-form multiplet $U$. The real auxiliary field ${\rm D}$ of the multiplet $V$ is here replaced by the real gauge three-form $C_{mnp}$.}
\label{tab:comp2}
\end{table}

The variation of the Lagrangian \eqref{VLagMaster} with respect to the vector multiplet $V$ produces the superspace equations of motion
\begin{equation}
\label{VsolLam}
\Lambda = \frac12\left(D^\alpha W_\alpha+\bar D_{\dot\alpha} \bar W^{\dot \alpha}\right) = D^\alpha W_\alpha\,,
\end{equation}
whose lowest component is
\begin{equation}
\Lambda| = 2\, {}^*\! G_4\,.
\end{equation}
Substituting \eqref{VsolV} and \eqref{VsolLam} in \eqref{VLagMaster} we get
\begin{equation}
\label{VLag3f}
\mathcal{L} = \left(\frac14 \int d^2\theta\, W^\alpha W_\alpha + \text{c.c.} \right)+ \mathcal{L}_{\rm bd}
\end{equation}
where the boundary terms are given by
\begin{equation}
\begin{split}
\label{VLagBd}
\mathcal{L}_{\rm bd} =&-\frac18 \int d^2\theta\,\bar D^2 (\Lambda V) -\frac18 \int d^2\bar\theta\, D^2 (\Lambda V) \\
&+\frac18 \left[ \int d^2\theta\, \bar D^2 (\Lambda \Sigma)+\int d^2\bar\theta\, D^2 (\Lambda \bar \Sigma)\right]\\
=&\, \frac{1}{64} [D^2,\bar D^2] \left(\Lambda (\bar \Sigma -  \Sigma)\right)|\,.
\end{split}
\end{equation}
In components, this Lagrangian is
\begin{equation}
\label{VLagOffSb}
\mathcal{L} =- \frac 14  F^{mn}F_{mn}  -\frac{1}{2\cdot 4!} G^{mnpq} G_{mnpq}  + \mathcal{L}_{\rm bd}\,,
\end{equation}
with
\begin{equation}
\label{VLagbdc}
\mathcal{L}_{\rm bd} = \frac{1}{3!}\partial_m \left(G^{mnpq} C_{npq}\right)\,.
\end{equation}
Setting the gauge three-form on-shell we immediately get
\begin{equation}	
G_{mnpq} = \frac{\xi}{2} \varepsilon_{mnpq}
\end{equation}
with $\xi$ a real constant. Plugging this solution back into \eqref{VLagOffSb}, we obtain the on-shell Lagrangian \eqref{VLagOnSa}. We conclude then that the role of the gauge three-form in the parent theory \eqref{VLagOffSb} is to dynamically generate the Fayet--Iliopoulos parameter $\xi$ as expectation value of ${}^*\!G_4$. As for the previous case, this dynamically generated parameter is related to the scale of supersymmetry breaking.

To sum up, we have shown how it is possible to reformulate generic Lagrangians involving chiral and vector superfields by means of only one ingredient, that is a complex linear superfield $\Sigma$. In the off-shell formulation of these Lagrangians, the non-propagating degrees of freedom are encoded into gauge three-forms and, as a consequence, the parameters contributing to the breaking of supersymmetry are generated dynamically as vacuum expectation values. In the following we extend the discussion to the case of $\mathcal{N}=2$ supersymmetry.


\section{$\mathcal{N}=2$ supersymmetry}

\label{sec:N2Formalism}

Before examining how gauge three-forms can be embedded into rigid $\mathcal{N}=2$ supersymmetric theories, we briefly review known facts about the construction of $\mathcal{N}=2$ Lagrangians within the superspace approach. We are interested in particular in how to rephrase the expressions in the language of $\mathcal{N}=1$ superspace.

\subsection{$\mathcal{N}=2$ chiral and vector multiplets}

\label{sec:N2FormalismMult}

The basic bricks which we shall need in the next section to build $\mathcal{N}=2$ Lagrangians are the chiral and vector (or reduced chiral) multiplets. They will be defined in an $\mathcal{N}=2$ superspace equipped, along with the space-time coordinates, with two sets of fermionic coordinates $\theta^\alpha$ and $\tilde \theta^\alpha$ associated to the two supersymmetry generators $Q^\alpha$ and $\tilde Q^\alpha$. The algebra satisfied by the $\mathcal{N}=2$ superspace derivatives without central charges is
\begin{equation}
\begin{aligned}
&\{ D_\alpha, \bar D_{\dot \alpha} \} = \{ \tilde D_\alpha, \bar{\tilde D}_{\dot \alpha}\} = -2i \sigma^m_{\alpha \dot \alpha}\partial_m,\\
&\{D_\alpha, D_\beta\} = \{\tilde D_\alpha, \tilde D_\beta\} = \{D_\alpha \tilde D_\beta\} = \{D_\alpha, \bar{\tilde D}_{\dot \beta}\}=0,
\end{aligned}
\end{equation}
where $\tilde D_\alpha$ generates the second supersymmetry.

The $\mathcal{N}=2$ chiral multiplet $\mathcal{A}$ can be represented by a superfield, which is chiral along the two fermionic directions
\begin{equation}
\label{N2Achirconst}
	\bar D_{\dot \alpha} \mathcal{A} = 0\,,\quad \bar {\tilde D}_{\dot \alpha} \mathcal{A} = 0\,.
\end{equation}
It has $16+16$ off-shell degrees of freedom, which are encoded in three $\mathcal{N}=1$ chiral superfields $X$, $\Phi$ and $W_\alpha$. 
It can be expanded in the $\tilde \theta$ coordinates as
\begin{equation}\label{N=2VecMult}
\begin{aligned}
\mathcal{A}(y,\theta, \tilde \theta) 
&= X(y,\theta) + \sqrt 2 \tilde\theta^\alpha W_\alpha(y,\theta) +\tilde \theta^2 \left(\frac{i}{2}\Phi(y,\theta)+\frac14 \bar D^2\bar X\right),
\end{aligned}
\end{equation}
where $y$ collects the chiral spacetime coordinates. The supersymmetry transformations of the $\mathcal{N}=1$ components of $\mathcal{A}$ along the $\tilde \theta$ coordinates are given by
\begin{align}
\tilde\delta X &= \sqrt 2 \eta^\alpha W_\alpha\,,\\[2mm]
\tilde \delta W &= i\sqrt 2 \sigma^m\bar\eta \partial_m X + \sqrt 2 \eta \left(\frac i2 \Phi+\frac14 \bar D^2 \bar X\right)\,,\\[2mm]
\tilde \delta \Phi  &= 2\sqrt 2 i \left(\frac14 \bar D^2(\bar\eta\bar W)-i\bar\eta\bar\sigma^m\partial_m W\right).
\end{align}
We stress that here the chiral superfield $W_\alpha$ does not satisfy any Bianchi identities and it cannot be explicitly written in terms of an $\mathcal{N}=1$ real potential. In other words, $W_\alpha$ does not represent the usual field strength of an $\mathcal{N}=1$ vector multiplet. 
The auxiliary components of $\mathcal{A}$ are defined as the projections
\begin{equation}
\begin{aligned}
-\frac14 D^2\mathcal{A} | &= f\,,
\\
-\frac14 \tilde D^2\mathcal{A} | &= \frac{i}{2} \varphi -\bar f\,,
\\
-\frac14  D \tilde D\mathcal{A} |&= \frac{{\rm D}}{\sqrt{2}}\,.
\end{aligned}
\end{equation}

It is possible to rephrase this construction in a manifestly $\rm SU(2)_R$-covariant manner. We first collect the superspace coordinates $\theta$ and $\tilde \theta$ into a $\rm SU(2)_R$ doublet 
\begin{equation}
\theta_i =\left( 
\begin{array}{c}
\theta_\alpha\\
\tilde \theta_\alpha
\end{array}
\right)
\end{equation}
and we define the superspace derivatives $D^{ij} = D^{i\,\alpha} D^j_{\alpha}$.
The superspace expansion of $\mathcal{A}$ is then
\begin{equation}
\mathcal{A} = X + \sqrt 2 \theta_i \lambda^i + \theta_{i} \theta_j \mathbb{Y}^{ij}+\dots,
\end{equation}
where $\lambda^i$ is the $\rm SU(2)_R$ doublet containing the fermions, while $\mathbb{Y}^{ij} = -\frac14 D^{ij}\mathcal{A}|$ is a matrix containing the auxiliary fields
\begin{equation}
	 \mathbb{Y} = \begin{pmatrix}f &  \frac{{\rm D}}{\sqrt{2}} \\ \frac{{\rm D}}{\sqrt{2}} & \frac{i}{2} \varphi -\bar f \end{pmatrix},
\end{equation}
which defines an $\rm SU(2)_R$ triplet $\vec{Y}$ as
\begin{equation}
-2i \mathbb{Y} \equiv (\vec{\sigma} \cdot \vec{Y}) \sigma^2\,.
\end{equation}
More explicitly, $\vec{Y}$ reads
\begin{equation}
\label{TripletGen}
\vec{Y} = \begin{pmatrix}
	2\, \text{Im} f  + \frac{\varphi}{2}\\ - 2\,\text{Re} f+ \frac{i \varphi}{2} \\ \sqrt{2} D
\end{pmatrix} \,,
\end{equation}
whose entries are generically complex.

The $\mathcal{N}=2$ vector (or reduced chiral) superfield $\mathcal{A}_D$ can be obtained from the chiral multiplet \eqref{N=2VecMult} by imposing the constraint given in \cite{Grimm:1977xp, deRoo:1980mm}, which results in the reduction of its off-shell degrees of freedom to $8+8$. This is equivalent to set $\Phi=0$ directly in the superspace expansion \eqref{N=2VecMult} \cite{Ambrosetti:2009za}, giving
\begin{equation}\label{ChiralC}
\begin{split}
\mathcal{A}_D(y,\theta, \tilde \theta) &= X_D(y,\theta) + \sqrt 2 \tilde\theta^\alpha W_{D\alpha}(y,\theta) + \frac14\tilde \theta^2  \bar D^2\bar X_D
\end{split}
\end{equation}
and by requiring also that the condition is preserved by the second supersymmetry, $\tilde \delta\, \Phi=0$, for consistency. From this requirement one gets the Bianchi identities
\begin{equation}
D^\alpha W_{D\,\alpha} = \bar D_{\dot \alpha} \bar W^{\dot \alpha}_D\,,
\end{equation}
which imply that $W_D$ can be expressed as the field strength of a gauge potential real superfield $V_D$
\begin{equation}
W_{D\,\alpha} = -\frac14 \bar D^2 D_\alpha V_D\,.
\end{equation}
We also stress that, for a reduced multiplet, the auxiliary field triplet $\vec{Y}$ of \eqref{TripletGen} is real. Adopting a manifestly $\rm SU(2)_R$ invariant notation, in which we define the doublet of fermions
\begin{equation}
{\bm \Psi} = \left(
\begin{array}{c}
\lambda\\
\psi
\end{array}
\right)
\end{equation}
and the supersymmetry parameters
\begin{equation}
{\bm \eta} = \left(
\begin{array}{c}
\eta_1\\
\eta_2
\end{array}
\right),
\end{equation}
the supersymmetry transformations of the vector multiplet can be written as
\begin{subequations}
\begin{align}
\delta \varphi &= \sqrt{2} {\bm\eta} {\bm\Psi}\,, \label{Susy2VarPhi}
\\
\delta v^m &= i ({\bm\eta} \sigma^m \bar {\bm\Psi}  + \bar {\bm \eta} \bar \sigma^m {\bm \Psi})\,, \label{Susy2VarB}
\\
\delta {\bm \Psi} &= i \sqrt{2} \sigma^m \bar {\bm \eta} \partial_m \varphi + \sigma^{mn} {\bm \eta}
 F_{mn}+ \frac{i}{\sqrt{2}} (\vec{\sigma} \cdot \vec{Y}) {\bm \eta}\,, \label{Susy2VarPsi}
\\
\delta \vec{Y} &= \sqrt{2}\, \bar {\bm \eta} \vec{\sigma}^m\, \partial_m {\bm \Psi} + \text{h.c.}\,. \label{Susy2VarY}
\end{align}
\end{subequations}

\subsection{Structure of the $\mathcal{N}=2$ Lagrangian}

\label{sec:N2Lag}

In this section we review the structure and the properties of $\mathcal{N}=2$ supersymmetric Lagrangians for an arbitrary number of vector multiplets.

For simplicity, let us start by considering the case of a single vector multiplet \eqref{ChiralC}. We define a holomorphic, but otherwise general, prepotential $F(\mathcal{A}_D)$, in terms of which a manifestly $\mathcal{N}=2$ Lagrangian can be built as the integral over the chiral $\mathcal{N}=2$ superspace
\begin{equation}
\label{LN=2b}
\mathcal{L} =\frac i2 \int d^2\theta d^2\tilde \theta F(\mathcal{A}_D)+{\rm c.c.}\,.
\end{equation}

By using the expansion \eqref{ChiralC} and integrating over the fermionic coordinates $\tilde \theta$, the Lagrangian \eqref{LN=2b} can be recast in the more familiar $\mathcal{N}=1$  language as
\begin{equation}
\label{LN=1}
\mathcal{L} = \left(\frac14 \int d^2\theta \tau(X) W^\alpha W_\alpha + {\rm c.c.}\right) +\int d^4\theta K(X,\bar X)\,,
\end{equation}
where 
\begin{equation}
\tau(X) =- i F_{XX}, \qquad K(X,\bar X) = \frac{i}{2}\left(X \bar F_{\bar X} - \bar X F_X \right)\,
\end{equation}
and $\partial_X \partial_{\bar X}K = \text{Im}\, F_{XX}$ is the metric of the special K\"ahler scalar manifold.
This Lagrangian has been written in a so called electric frame, in which only electric vector fields are present. Alternatively, with an $\rm SL(2,\mathbb{R})$ electro-magnetic duality transformation, it is possible to bring it into a magnetic frame, in which the electric vectors are exchanged with their magnetic dual.  In fact, let us consider, rather than \eqref{LN=2b}, the Lagrangian
\begin{equation}
\label{LN=2}
\mathcal{L} = \frac{i}{2} \int d^2 \theta d^2 \tilde \theta \left[F( \mathcal{A}) -  \mathcal{A}_D  \mathcal{A} \right] + {\rm c.c.}\,,
\end{equation}
where $\mathcal{A}$ is a chiral multiplet, while $\mathcal{A}_D$ is the magnetic dual of a vector multiplet. By integrating out $\mathcal{A}$, the Lagrangian is expressed entirely in the magnetic frame while, by integrating out $\mathcal{A}_D$, the constraint on $\mathcal{A}$ is imposed which reduces it to an $\mathcal{N}=2$ vector multiplet. In this sense, $\mathcal{A}_D$ can be thought of as a Lagrange multiplier. This can be most readily seen by rewriting \eqref{LN=2} in the language of $\mathcal{N}=1$ superspace
\begin{equation}
\label{LN=1b}
\begin{aligned}
\mathcal{L} &= -\frac i4 \int d^2\theta \left( F_{XX}  W^\alpha W_\alpha - 2 W_D^\alpha W_\alpha \right)\\
& - \frac i2 \int d^4 \theta F_X \bar{X} + \frac 14 \int d^2 \theta \Phi \left( X_D - F_X\right) + {\rm c.c.}\,.
\end{aligned}
\end{equation}
The equations of motion of the $\mathcal{N}=1$ superfields $X_D$ and $V_D$ contained in $\mathcal{A}_D$ give, respectively, the Bianchi identities of $W_\alpha$ and the constraint which reduces $\mathcal{A}$ to a $\mathcal{N}=2$ vector multiplet:
\begin{align}
\label{deltaVD} \delta V_D: &\qquad D^\alpha W_\alpha = \bar D_{\dot \alpha} \bar W^{\dot \alpha} \,,\\
\label{deltaXD} \delta X_D: &\qquad \Phi=0\,.
\end{align}
We note that we cannot vary with respect to $W_D$, being it constrained by the Bianchi identities, but it is indeed necessary to vary with respect to the real, unconstrained potential $V_D$. Hence, we immediately recognize that the integration of the constrained superfield $\mathcal{A}_D$ has the role to set the constraints on $\mathcal{A}$, so that \eqref{LN=1b} is identified with \eqref{LN=2b}.

As proposed in \cite{Antoniadis:1995vb}, however, a third possibility is to work in a frame which contains both electric and magnetic vectors at the same time. In fact the Lagrangian \eqref{LN=2} can be supplemented with both electric and magnetic Fayet--Iliopoulos parameters. Introducing the complex $\vec{E}$ and the real $\vec{M}$ parameters, we can add new couplings linear in $\vec{Y}$ and $\vec{Y}_D$ to obtain
\begin{equation}
\label{N2Lag}
\mathcal{L} = \frac{i}{2} \int d^2 \theta d^2 \tilde \theta \left[F( \mathcal{A}) - \mathcal{A}_D \mathcal{A} \right] + \frac12 \left(\vec{E}\cdot \vec{Y}+  \vec{M}\cdot \vec{Y}_D\right)+\text{c.c.}\,.
\end{equation}
These correspond respectively to electric and magnetic abelian gaugings of the theory and, but for those appearing possibly in the prepotential, they are the only parameters which are compatible with $\mathcal{N}=2$ supersymmetry.
In this sense, therefore, the case of extended supersymmetry is more constrained with respect to the $\mathcal{N}=1$ situation, in which a large class of parameters can enter the superpotential. In addition, in order to preserve the R-symmetry, we assume that $\vec{E}$ and $\vec{M}$ transform as triplets  under $\rm SU(2)_R$.
We note that the auxiliary fields $\vec{Y}_D$ transform as total derivatives under the supersymmetry transformation, in contrast to $\vec{Y}$, which transform as total derivative only once the superfield $\mathcal{A}_D$ is integrated out. The integration of $X_D$ gives indeed
\begin{equation}
\delta X_D: \qquad \Phi = 4 (M^2 + i M^1)\,,
\end{equation}
which means that $\Phi$ is a constant superfield. For consistency, the condition $\tilde \delta\, \Phi=0$ has to be imposed again.

The Lagrangian \eqref{N2Lag} can be easily generalized to the case of an arbitrary number of self-interacting chiral multiplets $\mathcal{A}^\Lambda$, with $\Lambda,\Sigma,\ldots=1,\ldots, N$, accompanied with an equal number of vector multiplets $\mathcal{A}_{D\,\Lambda}$ setting the constraints on $\mathcal{A}^\Lambda$:
\begin{equation}
\label{N2NLag}
\mathcal{L} = \frac{i}{2} \int d^2 \theta d^2 \tilde \theta \left[F( \mathcal{A}) - \mathcal{A}_{D\,\Lambda} \mathcal{A}^\Lambda \right] + \frac12 \left(\vec{E}_\Lambda\cdot \vec{Y}^\Lambda+  \vec{M}^\Lambda\cdot \vec{Y}_{D \Lambda}\right)+\text{c.c.}\,.
\end{equation}
After the Lagrange multiplier $\mathcal{A}_D$ is integrated out, the equations of motion of the auxiliary fields give
\begin{equation}
\vec{Y}^\Lambda = - 2\, \mathcal{N}^{\Lambda \Sigma}\left(\text{Re}\, \vec{E}_\Sigma + \text{Re}\, F_{\Sigma \Gamma}\vec{M}^\Gamma\right)+2i\vec{M}^\Lambda\,,
\end{equation}
where we have defined the metric of the special K\"ahler scalar manifold $\mathcal{N}_{\Lambda\Sigma} = \text{Im}\, F_{\Lambda\Sigma}$, together with its inverse $\mathcal{N}^{\Lambda\Sigma}=({\mathcal{N}_{\Lambda\Sigma}})^{-1}$. Substituting this expression for the auxiliary fields back into the Lagrangian, the following scalar potential is produced
\begin{equation}
\label{VpotN=2}
\mathcal{V} = \mathcal{N}^{\Lambda \Sigma}\left(\text{Re}\, \vec{E}_\Lambda + F_{\Lambda\Gamma}\vec{M}^\Gamma\right)\cdot\left(\text{Re}\, \vec{E}_\Sigma + \bar F_{\Sigma\Delta}\vec{M}^\Delta\right)+2 \text{Im}\, \vec{E}_\Lambda \cdot \vec{M}^\Lambda\,.
\end{equation}
Notice that $9N$ real parameters are appearing in the scalar potential. However $3N$ of them, namely those encoded in $\text{Im}\, \vec{E}_\Lambda$, are contributing solely as an additive constant to the Lagrangian and thus they can be disregarded, as long as we are focusing only on rigid supersymmetric theories.

In the following sections, by using variant $\mathcal{N}=1$ multiplets containing gauge three-forms, we will be able to rephrase the Lagrangians \eqref{N2Lag} and \eqref{N2NLag} in terms of peculiar $\mathcal{N}=2$ vector multiplets which dynamically generate the gauging parameters $\vec{E}$ and $\vec{M}$.


\section{Three-forms in $\mathcal{N}=2$ global supersymmetry}
\label{sec:N23form}

We extend now the procedure introduced in \cite{Farakos:2017jme} and reviewed in Section \ref{sec:N1case} to $\mathcal{N}=2$ Lagrangians of the kind of \eqref{N2Lag}. As in Section \ref{sec:N1case}, the Lagrangian \eqref{N2Lag} will be traded for an alternative one, which contains gauge three-forms and where the gauging parameters appear only when these gauge three-forms are set on-shell.
We here examine the case where the $\mathcal{N}=2$ Lagrangian is built out of a single abelian vector multiplet. The generalization to an arbitrary number $N$ of vector superfields is reported in the Appendix \eqref{app:Nvec}.

\subsection{The case of a single vector multiplet}
\label{sec:N23form1}

Let us consider the case of a single vector multiplet $\mathcal{A}$, whose $\mathcal{N}=2$ Lagrangian is \eqref{N2Lag}, endowed with Fayet--Iliopoulos parameters $\vec{E}$ and $\vec{M}$. In order to recast this Lagrangian in an $\mathcal{N}=1$ form, which is convenient for an analysis similar to that carried in Section \ref{sec:N1case}, we use the $\rm SU(2)$ R-symmetry of the theory and rotate the parameters such that
\begin{equation}
\label{N2LagGauging}
\Re \vec{E} =\left(0,-e,\frac{\xi}{2\sqrt{2}}\right)\,, \qquad \vec{M} = (0,-m,0)\,,
\end{equation}
with $e$, $m$ and $\xi$ real constants. Indeed, the $\rm SU(2)_R$ covariance of the Lagrangian \eqref{N2LagGauging} ensures that there is no loss of generality in the choice \eqref{N2LagGauging}. We have also discarded the imaginary part of $\vec{E}$, since we have already shown that it contributes only as an additive constant to the theory.

The Lagrangian \eqref{N2Lag}, after integrating out the constrained superfield $\mathcal{A}_D$, can be written in $\mathcal{N}=1$ language as
\begin{equation}\label{N2LagB}
\begin{aligned}
\mathcal{L} =&  \int d^4\theta\, K(X,\bar X)+ \left(\frac14 \int d^2\theta\, \tau (X) W^{\alpha} W_\alpha +\text{c.c.}\right)+
\\
& + \left( \int d^2\theta\,W(X)+\text{c.c.}\right) + \xi \int d^4\theta\, V\,,
\end{aligned}
\end{equation}
with the superpotential $W(X)$ given by
\begin{equation}
\label{N2W}
W(X)= e X + m F_X(X)\,.
\end{equation}
The bosonic components of \eqref{N2LagB} are
\begin{equation}
\label{N2LScom}
\begin{split}
\mathcal{L}\big|_{\text{bos}} =& -\text{Im}\, F_{XX}\, \partial_m \varphi\, \partial^m \bar \varphi -\frac 14 \text{Im}\, F_{XX} F^{mn}F_{mn} - \frac18 \text{Re}\, F_{XX}\, \varepsilon_{klmn} F^{kl} F^{mn} + \\
&+  \text{Im}\, F_{XX} f \bar f  +\frac12\text{Im}\, F_{XX} {\rm D}^2 + (e +  m F_{XX}) f + (e + m \bar F_{XX}) \bar f + \frac{\xi}{2} {\rm D}\,
\end{split}
\end{equation}
and, integrating out the auxiliary fields $f$ and ${\rm D}$, we arrive at
\begin{equation}
\label{N2LScomOS}
\begin{split}
\mathcal{L}\big|_{\text{bos}} =& -\text{Im}\, F_{XX}\, \partial_m \varphi\, \partial^m \bar \varphi -\frac 14 \text{Im}\, F_{XX} F^{mn}F_{mn}  - \frac18  \text{Re}\, F_{XX}\, \varepsilon_{klmn} F^{kl} F^{mn} - \mathcal{V} (\varphi,\bar\varphi) \,,	
\end{split}
\end{equation}
with the scalar potential
\begin{equation}
\label{N2LV}
\begin{split}
\mathcal{V} (\varphi,\bar\varphi) = \frac{1}{ \text{Im}\, F_{XX}} |e+m F_{XX}|^2 + \frac{\xi^2}{8\,  \text{Im}\,F_{XX}}\,.
\end{split}
\end{equation}

We would like to generate dynamically the gauging parameters $e$, $m$ and $\xi$ entering the scalar potential and the superpotential. To this purpose, we shall perform a two steps procedure.
First, we will trade the vector multiplet $\mathcal{A}_D$ in \eqref{N2Lag} for its variant version
\begin{equation}\label{ChiralS}
\begin{split}
\mathcal{S}_D(y,\theta, \tilde \theta) &= S_D (y,\theta) + \sqrt 2 \tilde\theta^\alpha W_{D\,\alpha}(y,\theta) + \frac14 \tilde \theta^2 \bar D^2\bar S_D\,,
\end{split}
\end{equation}
where $W_{D\,\alpha} =-\frac14 \bar D^2 D_\alpha U_D$. Here, the ordinary chiral multiplet $X$ and vector multiplet $V$ of  \eqref{ChiralC} are replaced with \eqref{N1solPhi} and \eqref{VsolV} respectively.
On the one hand, this will allow for promoting the magnetic parameters $\vec{M}$ to be dynamical; on the other, it will allow for establishing an off-shell correspondence between the $\mathcal{N}=2$ multiplets \eqref{N=2VecMult} and \eqref{ChiralS}. Then, after integrating out the variant $\mathcal{S}_D$ multiplet, we will proceed to the second step. It consists in an additional trading, which exchanges the remaining $\mathcal{A}$ multiplet with another variant chiral multiplet of the kind of \eqref{ChiralS}. The final Lagrangian, setting the residual three-forms on-shell, will coincide with the on-shell Lagrangian \eqref{N2LScomOS}.

\subsubsection*{First step: generating the magnetic parameters}
We start recalling that, in $\mathcal{N}=1$ language, \eqref{N2Lag} reads
\begin{equation}
\label{Alt_LN=1a}
\begin{aligned}
\mathcal{L} &= -\frac i4 \int d^2\theta \left( F_{XX}  W^\alpha W_\alpha - 2 W_D^\alpha W_\alpha \right)\\
&\quad - \frac i2 \int d^4 \theta F_X \bar{X} + \frac 14 \int d^2 \theta \Phi \left( X_D - F_X\right) + \frac12 \left(\vec{E}\cdot \vec{Y}+  \vec{M}\cdot \vec{Y}_D\right)+ {\rm c.c.}\,.
\end{aligned}
\end{equation}
In order to reinterpret the magnetic gauging parameters as originating from vacuum expectation values of gauge three-forms, we first promote the coupling $\vec{M}\cdot \vec{Y}_D$ to a full dynamical entity. 
Hence we rewrite \eqref{Alt_LN=1a} as
\begin{equation}
\label{Alt_LN=1D}
\begin{aligned}
\mathcal{L} &= \bigg\{-\frac i4 \int d^2\theta \left( F_{XX}  W^\alpha W_\alpha - 2 W_D^\alpha W_\alpha \right)
\\
& \quad\quad- \frac i2 \int d^4 \theta F_X \bar{X} + \frac 14 \int d^2 \theta \Phi \left( X_D - F_X\right) + \frac12 \vec{E}\cdot \vec{Y}+ {\rm c.c.}\bigg\}
\\
&\quad+\left\{\int d^2 \theta\, \left(\Lambda_1^D X_D + \frac14   \bar D^2 (\Sigma_{1D} \bar \Lambda_1^D) \right)+ {\rm c.c.}\right\}
\\
&\quad+\left\{\frac18 \int d^2\theta\,\bar D^2 [\Lambda_2^D (\Sigma_{2D} - V_D)] + \text{c.c.} \right\},
\end{aligned}
\end{equation}
As explained in Section \ref{sec:N1caseC} (see \eqref{N1LB}), the third line provides the gauge three-form inside the $\mathcal{N}=1$ chiral superfield in $\mathcal{A}_D$, while the fourth, as in \eqref{sec:N1caseV} (see \eqref{VLagMaster}), provides it for the vector superfield. 
In particular, $\Lambda_1^D$ and $\Lambda_2^D$ are respectively a chiral and a real superfield which play the role of Lagrange multipliers, while $\Sigma_1^D$ and $\Sigma_2^D$ are complex linear multiplets containing the gauge three-forms. As a consistency check, from the equations of motion of $\Sigma_{1D}$ and $\Sigma_{2D}$ we get
\begin{equation}
\label{Alt_LambdaOS}
	\Lambda_1^D = -M^2-i M^1\,,\qquad \Lambda_2^D= 2\sqrt{2} M^3\,,
\end{equation}
with $\vec{M}$ arbitrary real integration constants, recovering \eqref{Alt_LN=1a}. 

On the contrary, in order to obtain an $\mathcal{N}=2$ Lagrangian which contains gauge three-forms in place of the auxiliary fields $\vec{Y}_D$, we have to integrate out the chiral superfield $X_D$ and vector superfield $V_D$, as well as the Lagrange multipliers $\Lambda_1^D$ and $\Lambda_2^D$. 
The variations with respect to the Lagrange multipliers $\Lambda_1^D$ and $\Lambda^D_2$ give the relations
\begin{align}
\delta \Lambda_1^D: \quad X_D &= -\frac14 \bar D^2 \bar\Sigma_{1D} \equiv S_D\,, \label{Alt_XDsol}
\\
\delta \Lambda_2^D: \quad  V_D &=  \frac{\Sigma_{2D} + \bar \Sigma_{2D}}{2} \equiv U_D\,, \label{Alt_VDsol}
\end{align}
which trade, respectively, the ordinary $\mathcal{N}=1$ chiral multiplet $X_D$ and vector multiplet $W_D^\alpha$ for a double and a vector three-form multiplet. Indeed, the variations with respect to the ordinary $\mathcal{N}=1$ superfields $X_D$ and $V_D$ give
\begin{align}
\label{Alt_L1Dsol}
\delta X_D: \quad \Lambda^D_1 &= -\frac 14 \Phi\,,
\\
\label{Alt_L2Dsol}
\delta V_D: \quad \Lambda^D_2 &= - {\rm Im} (D^\alpha W_\alpha)\,.
\end{align}
Plugging (\ref{Alt_XDsol}-\ref{Alt_L2Dsol}) in \eqref{Alt_LN=1D}, we get
\begin{equation}
\label{Alt_LN=1Db}
\begin{aligned}
\mathcal{L} &= \bigg\{-\frac i4 \int d^2\theta \left( F_{XX}  W^\alpha W_\alpha - 2 W_D^\alpha W_\alpha \right)
\\
& \quad\quad- \frac i2 \int d^4 \theta F_X \bar{X} + \frac 14 \int d^2 \theta \Phi \left( S_D - F_X\right) + \frac12 \vec{E}\cdot \vec{Y}+ {\rm c.c.}\bigg\} + \mathcal{L}_{\rm bd}^{(D)}
\end{aligned}
\end{equation}
with
\begin{equation}\label{Alt_LN=1Dbd}
\begin{split}
\mathcal{L}_{\rm bd}^{(D)}= &-\frac14 \left(\int d^2\theta \bar D^2 - \int d^2\bar\theta D^2\right) \left[\left(-\frac 14 \Phi  \right) \bar\Sigma_{1D}\right]\\
&+\frac{1}{16} \left(\int d^2\theta \bar D^2 - \int d^2\bar\theta D^2\right)  \left[ - \text{Im} (D^\alpha W_\alpha) \Sigma_{2D} \right] +\text{c.c.} \, .
\end{split}
\end{equation}

Before moving on and integrating out the superfields $S_D$ and $U_D$, let us notice that the Lagrangian \eqref{Alt_LN=1Db} can be recast in a manifest $\mathcal{N}=2$ form as
\begin{equation}
\label{Alt_LN=2}
\mathcal{L} = \frac{i}{2} \int d^2 \theta d^2 \tilde \theta \left[F( \mathcal{A}) -  \mathcal{S}_D  \mathcal{A} \right] +  \frac12 \vec{E}\cdot \vec{Y} + {\rm c.c.} + \mathcal{L}_{\rm bd}^{(D)}\,.
\end{equation}
As desired, we have promoted the usual reduced chiral Lagrange multiplier $\mathcal{A}_D$ appearing in \eqref{N2Lag} to $\mathcal{S}_D$, which is a multiplet of the variant type as in \eqref{ChiralS}.
It contains a chiral double three-form multiplet $S_D$ and a real vector three-form mulitplet $U_D$, whose non propagating degrees of freedom are given by 
\begin{equation}\label{Alt_Y3f}
 \begin{split}
 -\frac14 D^2\mathcal{S}_D | &= -i  {}^*\!\bar F_{4}^D = - \frac{i}{3!} \varepsilon^{mnpq} \partial_{m} \bar{\mathcal{B}}_{ npq}^D \,,
 \\
 -\frac14 \tilde D^2\mathcal{S}_D | &= -i  {}^*\!F_{4}^D = - \frac{i}{3!} \varepsilon^{mnpq} \partial_{m} \mathcal{B}_{ npq}^D \,,
 \\
 -\frac14  D \tilde D\mathcal{S}_D |&=-\frac{1}{\sqrt{2}} {}^*\!G_{4}^D = -\frac{1}{3! \sqrt{2}} \varepsilon^{mnpq} \partial_{m} C_{ npq}^D\,,
 \end{split}
 \end{equation}
with $\mathcal{B}^D_{mnp}$ and $C_{mnp}^D$ the components of a complex and a real gauge three-form respectively. 

Resuming the previous discussion, in order to retrieve a Lagrangian formulated solely in terms of the multiplet $\mathcal{A}$, the Lagrange multiplier $\mathcal{S}_D$ has to be integrated out. Owing to the presence of the gauge three-forms, in comparison with (\ref{deltaVD},\ref{deltaXD}), the variations with respect to $S_D$ and $U_D$ differ only in their auxiliary components. In fact, integrating out $C_{mnp}^D$ and  ${\mathcal{B}}_{ npq}^D$, we get
\begin{equation}
\label{Alt_DPhi_OS}
	{\rm Im}\, {\rm D} = \sqrt{2} M^3\, \qquad \Phi = 4 (M^2 +i M^1)
\end{equation}
consistently with \eqref{Alt_LambdaOS}, \eqref{Alt_L1Dsol} and \eqref{Alt_L2Dsol}. In other words, the choice of a variant Lagrange multiplier \eqref{ChiralS} results in a dynamical generation of the magnetic Fayet--Iliopoulos parameters $\vec{M}$ which appear in \eqref{N2Lag}. In order to make contact with \eqref{N2LagB} we choose $\vec{M} = (0,-m,0)$ and, as a consequence, the Lagrangian \eqref{Alt_LN=1Db} becomes
\begin{equation}
\label{Alt_LN=1c}
\begin{aligned}
\mathcal{L} &= 
-\frac i4 \int d^2\theta F_{XX}  W^\alpha W_\alpha  - \frac i2 \int d^4 \theta F_X \bar{X} + m \int d^2 \theta  F_X  + \frac12 \vec{E}\cdot \vec{Y}+ {\rm c.c.}.
\end{aligned}
\end{equation}
This Lagrangian, in comparison with \eqref{Alt_LN=1D} and recalling the component structure of the chiral multiplet \eqref{N=2VecMult}, suggests that integrating out $\mathcal{S}_D$ results in constraining $\mathcal{A}$ as
\begin{equation}\label{Alt_A}
\mathcal{A}(y,\theta, \tilde \theta) = X(y,\theta) + \sqrt 2 \tilde\theta^\alpha W_\alpha(y,\theta) +\tilde \theta^2 \left(-2i m+\frac14 \bar D^2\bar X\right).
\end{equation}
Due to the presence of gauge-three forms, therefore, a parameter related to a magnetic Fayet--Iliopoulos gauging has been inserted dynamically into the expression \eqref{ChiralC} of the $\mathcal{N}=2$ vector superfield. This parameter will play an important role when studying the mechanism of supersymmetry breaking, as shown in the next section.

\subsubsection*{Second Step: generating the electric parameters}
We have just presented a recipe in order to dynamically produce the magnetic gauging parameters, but for the moment nothing has be done on the electric gauging parameters $\vec{E}$.  If we insist on considering the multiplet $\mathcal{A}$ an ordinary one as in \eqref{N=2VecMult}, then the only choice is to add the electric gauging by hand from the start, as in \eqref{N2Lag}. However, if we relax this request there is still another option: we may assume that $\mathcal{A}$ could be a variant multiplet of the kind of \eqref{ChiralS}, which endows gauge three-forms in its non-propagating components, and allow for the dynamical generation of the electric gauging parameters as well. This represents the second step of the procedure we are proposing. We promote then \eqref{Alt_LN=1c} to
\begin{equation}\label{Alt_N2Master}
\begin{aligned}
\mathcal{L} =&  \int d^4\theta\, K(X,\bar X)+\left(\frac14 \int d^2\theta\, \tau(X) W^{\alpha} W_\alpha + \text{c.c.}\right)+\\
&+\left\{\int d^2\theta \Lambda_1 X
+\frac14 \int d^2\theta\, \bar D^2\left(\Sigma_1 \bar \Lambda_1\right)+ m \int d^2 \theta F_X+\text{c.c.}\right\}+\\
&+\left\{\frac18 \int d^2\theta\,\bar D^2 [\Lambda_2 (\Sigma_2 - V)] + \text{c.c.} \right\},
\end{aligned}
\end{equation}
The second line provides the exchange between the chiral multiplet $X$ and its three-form counterpart: $\Lambda_1$ is a chiral Lagrange multiplier and $\Sigma_1$ a complex linear multiplet. The third line trades the vector multiplet $W^\alpha$ for a vector three-form multiplet: here $\Lambda_2$ is a real Lagrange multiplier and $\Sigma_2$ a complex linear multiplet.
The Lagrangian \eqref{Alt_N2Master} truly reproduces \eqref{N2LagB}. This can be most readily seen from the integration of the complex linear superfields $\Sigma_1$ and $\Sigma_2$, which sets
\begin{equation}\label{Alt_Lambda_sol}
	\Lambda_1 = e\,, \qquad \Lambda_2 = \xi\,,
\end{equation}
with $e$ and $\xi$ arbitrary real constants. Plugging \eqref{Alt_Lambda_sol} in \eqref{Alt_N2Master},  we reobtain \eqref{N2LagB}, ensuring the equivalence of the two descriptions. In order to re-express instead \eqref{Alt_N2Master} in terms of the three-form multiplets, we have to integrate out both the Lagrange multipliers $\Lambda_1$, $\Lambda_2$ and the ordinary $\mathcal{N}=1$ superfields $X$ and $V$. The variations with respect to the Lagrange multipliers $\Lambda_1$ and $\Lambda_2$ give
\begin{align}
\delta \Lambda_1: \quad X &= -\frac14 \bar D^2 \bar\Sigma_1 \equiv S\,, \label{Alt_Xsol}\\
\delta \Lambda_2: \quad  V &=  \frac{\Sigma_2 + \bar \Sigma_2}{2} \equiv U\,, \label{Alt_Vsol}
\end{align}
which trade the $\mathcal{N}=1$ multiplets for their three-form counterparts, while the variations with respect to $X$ and $V$ produce
\begin{align}
\label{Alt_L1sol}
\delta X: \quad \Lambda_1 &= \frac 14 \bar{D}^2 K_{X} - m F_{XX}+ \frac{i}4 F_{XXX} W^\alpha W_\alpha\,,
\\
\label{Alt_L2sol}
\delta V: \quad \Lambda_2 &= \text{Im}\, F_{XX} D^\alpha W_\alpha + \frac12 \left[(D^\alpha \tau) W_\alpha - (\bar D_{\dot \alpha} \bar \tau) \bar W^{\dot \alpha} \right]\,.
\end{align}
Inserting (\ref{Alt_Xsol}--\ref{Alt_L2sol}) in \eqref{Alt_N2Master} we get
\begin{equation}
\label{Alt_N2LD}
\mathcal{L} = \int d^4\theta K(S,\bar S) + \left( \frac14 \int d^2\theta \tau(S) W^{\alpha} W_{\alpha}+ m \int d^2 \theta F_S+ \text{c.c.}\right) +\mathcal{L}_{\rm bd},
\end{equation} 
with the boundary terms
\begin{equation}\label{Alt_N2LDbd}
\begin{split}
\mathcal{L}_{\rm bd}= &-\frac14 \left(\int d^2\theta \bar D^2 - \int d^2\bar\theta D^2\right) \left[\left(\frac 14 \bar{D}^2 K_{S} - m F_{SS} + \frac{i}4 F_{SSS} W^\alpha W_\alpha  \right) \bar\Sigma_1\right]\\
&+\frac{1}{16} \left(\int d^2\theta \bar D^2 - \int d^2\bar\theta D^2\right)  \left[ \left(\text{Im}\, F_{SS} D^\alpha W_\alpha +\frac12 W^\alpha D_\alpha \tau  + \frac12  \bar W_{\dot \alpha} \bar D^{\dot \alpha} \bar \tau \right)  \Sigma_2 \right]
\\
& +\text{c.c.} \, .
\end{split}
\end{equation}
The bosonic components of \eqref{Alt_N2LD} are
\begin{equation}
\label{Alt_N2LScom}
\begin{split}
\mathcal{L}\big|_{\text{bos}} =& -\text{Im}\, F_{SS}\, \partial_m s\, \partial^m \bar s -\frac 14 \text{Im}\, F_{SS} F^{mn}F_{mn} - \frac18 \text{Re}\, F_{SS}\, \varepsilon_{klmn} F^{kl} F^{mn}  \\
&+ \text{Im}\, F_{SS} {}^*\! {F}_4 {}^*\!\bar{{F}}_4    +\frac12  \text{Im}\, F_{SS} ({}^*\!G_4)^2 +\left(-i m F_{SS} {}^*\!\bar{{F}}_4 + {\rm c.c.}\right)+ \mathcal{L}_{\rm bd}
\end{split}
\end{equation}
with
\begin{equation}\label{Alt_N2Lbdcom}
\begin{split}
\mathcal{L}_{\rm bd} = &\left\{\frac{1}{3!} \partial_k\left[ B_{lmn}\left(\text{Im}\, F_{SS}\bar F^{klmn}-i m\, \varepsilon^{klmn} \bar F_{SS}\right)\right]+\text{c.c.}\right\}\\
&+\frac{1}{3!}\partial_k \left(\text{Im}\, F_{SS}\, C_{lmn}  G^{klmn}\right) \, .
\end{split}
\end{equation}
By setting the gauge three-forms on shell as
\begin{align}
\label{Alt_Fos}
F_{klmn} &=-\frac{i}{{\rm Im}\, F_{SS}}(e+m F_{SS})\varepsilon_{klmn}\,,
\\
\label{Alt_Gos}
G_{klmn} &= -\frac{\xi}{2\, \text{Im}\, F_{SS}}\, \varepsilon_{klmn}\,,
\end{align}
the same potential as \eqref{N2LV} is recovered.

We notice finally that the Lagrangian \eqref{Alt_N2LD} can also be recast in $\mathcal{N}=2$ superspace as 
\begin{equation}
\label{N2LagSS}
\mathcal{L} = \frac{i}{2} \int d^2 \theta d^2 \tilde \theta F( \mathcal{S})  +\text{c.c.} + \mathcal{L}_{\text{bd}}\,.
\end{equation}
provided that we introduce the $\mathcal{N}=2$ multiplet defined as
\begin{equation}\label{Alt_S}
\mathcal{S}(y,\theta, \tilde \theta) = S(y,\theta) + \sqrt 2 \tilde\theta^\alpha W_\alpha(y,\theta) +\tilde \theta^2 \left(-2 i m+\frac14 \bar D^2\bar S\right).
\end{equation}
where $S$ is that introduced \eqref{Alt_Xsol} and $W_{\alpha} = -\frac14 \bar D^2 D_\alpha U$, with $U$ as in \eqref{Alt_Vsol}. This multiplet provides a direct off-shell correspondence with the chiral multiplets \eqref{N=2VecMult} or \eqref{Alt_A}.

\subsubsection*{Summary of the results}
In this section we have presented a two steps procedure which allows for a dynamical generations of the parameters entering the superpotential and the scalar potential of generic $\mathcal{N}=2$ globally supersymmetric models. The first step leads to a dynamical generation of the magnetic gauging parameters. Once they are generated by integrating out the corresponding gauge three-forms, we can proceed to a second step. This amounts to promote also the electric gauging parameters to expectation values of four-form field strengths. The final result is the off-shell Lagrangian \eqref{N2LagSS}. In other words, the inclusion of parameters in the original Lagrangian has been traded for a boundary condition problem. 
Indeed the two steps can be performed at the same time, by conveniently combining \eqref{Alt_LN=1D} and \eqref{Alt_N2Master}, but we preferred to keep them separate for the plainness of the discussion. 
In the next section we are going to apply the formalism developed so far to analyse the mechanism of spontaneous partial breaking of supersymmetry.

\subsection{An alternative procedure}

In the previous subsection we proposed a two steps procedure in order to dynamically generate parameters in generic $\mathcal{N}=2$ supersymmetric theories. In particular, the magnetic Fayet--Iliopoulos parameter $m$ is generated first, while its electric counterparts are introduced in a second step. Even though the steps can be performed at the same time, in the sense that they do not clash, they generate different parts of the superpotential and of the scalar potential. One can however wonder if it would be possible to generated dynamically these quantities solely in one passage, following a logic closely related to that of \cite{Farakos:2017jme}.
It turns out indeed that the procedure of \cite{Farakos:2017jme} can be applied directly to the $\mathcal{N}=2$ case and the entire superpotential can be generated dynamically in a single step, but only for a particular class of models.  
In the present subsection we depict this alternative procedure for a system with a generic number of vector superfields, the minimal case of one single superfield being somehow trivial, as it is going to be clear in a while. The reader interested in the analysis of the partial breaking of supersymmetry can skip this subsection at first.

We start from the Lagrangian
\begin{equation}
\label{N2NNMaster}
\begin{aligned}
\mathcal{L} =&  \int d^4\theta\, K(X,\bar X)+\left(\frac14 \int d^2\theta \tau_{\Lambda \Sigma}(X) W^{\Lambda\alpha} W^\Sigma_\alpha + c.c\right)\\
&+\left\{\int d^2\theta \Lambda_{1\Sigma} X^\Sigma-\frac14 \int d^2\theta\, \bar D^2\left[  \Sigma_{1\Lambda} \,\mathcal{N}^{\Lambda\Sigma}(\Lambda_{1\Sigma}-\bar \Lambda_{1\Sigma})\right]+\text{c.c.}\right\}\\
&+\left\{\frac18 \int d^2\theta\,\bar D^2 [\Lambda_{2\Gamma} (\Sigma^\Gamma_2 - V^\Gamma)] + \text{c.c.} \right\}\,,
\end{aligned}
\end{equation}
where the second line comes from trading the superpotential term in \eqref{N2LagB} for an expression introducing gauge three-forms dynamically: $\Lambda_{1\Sigma}$ are chiral superfields which play the role of Lagrange multipliers and $\Sigma_{1\Lambda}$ are complex linear multiplets. The third line, instead, is the dynamical promotion of the Fayet--Iliopoulos term: $\Sigma_{2}^\Gamma$ are complex linear multiplets while $\Lambda_{2\Gamma}$ are real Lagrange multipliers superfields.

The equations of motion for $\Sigma_{1\Lambda}$,
\begin{equation}
D_\alpha[\mathcal{N}^{\Lambda \Sigma}(\Lambda_{1\Sigma}- \bar \Lambda_{1\Sigma})]=0,
\end{equation}
impose indeed that $\Lambda_{1\Sigma} = e_\sigma + m^\Gamma F_{\Sigma \Gamma}$.
On the other hand, integrating out the complex linear superfield $\Sigma_2$, we immediately get that $\Lambda_{2\Sigma}$ are just real constants $\xi_{\Sigma}$. Plugging these back into the Lagrangian \eqref{N2NNMaster}, we must obtain again \eqref{N2LagB}. It is however immediate to realize that, for this to happen, we have to require the prepotential $F(\mathcal{A})$ to be a degree--two homogeneous function of its argument, in order that $F_{\Lambda\Sigma}X^\Sigma = F_\Lambda$ and the superpotential \eqref{N2W} is recovered. This restriction is avoided in the two steps procedure we presented before.\footnote{We notice that, in the case of a single vector superfield, the homogeneity fixes $F(\mathcal{A})=\frac i2 \mathcal{A}^2$. Its second derivative is then constant and part of the following discussion becomes trivial.}

Despite this fact, we can still produce a Lagrangian which contains gauge three-forms in place of the auxiliary fields $\vec{Y}^\Lambda$. To this purpose, we have to integrate out the chiral superfields $X^\Lambda$ and the vector superfields $V^\Lambda$, as well as the Lagrange multipliers $\Lambda_{1\Sigma}$ and $\Lambda_{2\Sigma}$. The variation with respect to the Lagrange multipliers $\Lambda_{1\Sigma}$ re-expresses the chiral superfields $X^\Lambda$ as double three-form multiplets as
\begin{equation}
\label{N2Xsoll}
\begin{aligned}
X^\Lambda &= \frac14 \bar D^2 \left[(N^{\Lambda \Gamma} (\Sigma_{1\Gamma}-\bar \Sigma_{1\Gamma})\right]\equiv S^\Lambda\,.
\end{aligned}
\end{equation}
Notice that this relationship is non-linear in $S^\Lambda$ and it might not be possible in general to solve it and obtain an explicit expression for these superfields, but we can nevertheless understand their main properties.
We can calculate first of all the lowest components
\begin{equation}
S^\Lambda | = \mathcal{N}^{\Lambda \Gamma} \left(-\frac14 \bar D^2 \bar \Sigma_{1\,\Gamma} \big|\right)+\text{fermions} \equiv s^\Lambda +\text{fermions}\,.
\end{equation}
We notice then that \eqref{N2Xsoll} is left invariant by the gauge transformations
\begin{equation}
\Sigma_{1\, \Gamma} \to \Sigma_{1\, \Gamma} + \tilde L_\Gamma+\bar F_{\Gamma \Delta} L^\Delta,
\end{equation}
where $\tilde L_\Lambda$ and $ L^\Lambda$ are real linear superfields. In order to be compatible with this gauge transformation, the gauge three-forms in $S^\Lambda$ have to appear in
\begin{equation}
-\frac14 D^2 S^\Lambda | = -i \mathcal{N}^{\Lambda \Sigma}\, {}^*\!\mathcal{F}_{4\Sigma} + \text{fermions}\,,
\end{equation}
within the specific combination of four-forms
\begin{equation}
\mathcal{F}_{\Lambda\, klmn} \equiv \tilde F_{\Lambda \,klmn}+\bar F_{\Lambda \Sigma} F^\Sigma_{klmn}\,.
\end{equation}
In other words, in order to preserve the associated gauge invariance, which is necessary for the matching of the degrees of freedom, the complex gauge-three forms in $S^\Lambda$ are divided into two real parts, ${}^*\!F^\Lambda_4$ and ${}^*\!\tilde F_{4\,\Lambda}$ which appear in the Lagrangian combined inside ${}^*\!\mathcal{F}_{4\,\Lambda}$. This is more involved with respect to the analogous case in the previous subsection, but it is essential in order to reconstruct the correct on-shell form of the Lagrangian.

The variation with respect to the real Lagrange multipliers $\Lambda_{2\Sigma}$, instead exchanges the usual vector multiplets $V^\Lambda$ with their variant versions \eqref{Alt_Vsol}, as in the previous discussion.
The variations with respect to the chiral superfields $X^\Lambda$ and the vector superfields $V^\Lambda$ give respectively
\begin{equation}
\label{N2L1soll}
\begin{aligned}
\Lambda_{1\, \Sigma}&= \frac 14 \bar{D}^2 K_{\Sigma} - \frac14 \tau_{\Sigma \Gamma\Delta} W^{\Gamma\, \alpha} W^\Delta_{\alpha} \\
&- \frac18 \bar D^2 \left[(\Sigma_{1\,\Pi}-\bar \Sigma_{1\,\Pi}) (\Lambda_{1\,\Gamma}-\bar \Lambda_{1\,\Gamma}) \mathcal{N}^{\Pi\Lambda} \mathcal{N}^{\Delta\Gamma} \tau_{\Sigma \Lambda\Delta}\right]\,,\\
\Lambda_{2\,\Sigma}&= \text{Re}\, \tau_{\Sigma\Gamma} D^{\alpha} W^{\Gamma}_\alpha + \frac12 \left( D^\alpha \tau_{\Sigma \Gamma} W^{\Gamma}_ \alpha + \bar D_{\dot \alpha} \bar \tau_{\Sigma \Gamma} \bar W^{\Gamma \dot \alpha} \right)\,.
\end{aligned}
\end{equation}
Substituting \eqref{N2Xsoll}, \eqref{Alt_Vsol} and \eqref{N2L1soll} into \eqref{N2NNMaster} we obtain a new Lagrangian fully re-expressed in terms of the three-form multiplets
\begin{equation}
\label{N2LDD}
\mathcal{L} =  \int d^4\theta\, K(X,\bar X)+\left(\frac14 \int d^2\theta \tau_{\Lambda\Sigma}(X) W^{\Lambda\,\alpha} W^\Sigma_\alpha + c.c\right)+\mathcal{L}_{\rm bd},
\end{equation} 
with 
\begin{equation}\label{N2LDbdd}
\begin{split}
\mathcal{L}_{\rm bd}= &-\frac14 \left(\int d^2\theta \bar D^2 - \int d^2\bar\theta D^2\right) \left[\Lambda_{1\,\Sigma} \mathcal{N}^{\Sigma\Gamma} \bar\Sigma_{1\,\Gamma}\right]\\
&+\frac{1}{16} \left(\int d^2\theta \bar D^2 - \int d^2\bar\theta D^2\right)  \left( \Lambda_{2\,\Gamma}  \Sigma^\Gamma_2\right)	+\text{c.c.} \,.
\end{split}
\end{equation}
Its bosonic components are
\begin{equation}
\label{N2NLScomm}
\begin{split}
\mathcal{L}\big|_{\text{bos}} =& -\mathcal{N}_{\Lambda\Sigma}\, \partial_m s^\Lambda\, \partial^m \bar s^\Sigma -\frac 14 \mathcal{N}_{\Lambda\Sigma} F^{\Lambda\, mn}F^\Sigma_{mn}  - \frac18 \text{Re}\, F_{\Lambda\Sigma}\, \varepsilon_{klmn} F^{\Lambda\, kl} F^{\Sigma mn} \\
&+ \mathcal{N}^{\Lambda\Sigma} {}^*\!\mathcal{F}_{4\,\Lambda} {}^*\!\bar{\mathcal{F}}_{4\,\Sigma}    +\frac12  \mathcal{N}_{\Lambda\Sigma} {}^*\!G_4^\Lambda {}^*\!G_4^\Sigma+ \mathcal{L}_{\rm bd}
\end{split}
\end{equation}
with
\begin{equation}\label{N2Lbdcom}
\begin{split}
\mathcal{L}_{\rm bd} = &\left\{-\frac{1}{3!} \partial_k \left[\varepsilon^{klmn} \mathcal{N}^{\Lambda\Sigma} \left(\tilde A_{\Lambda \,lmn} + \bar F_{\Lambda\Sigma\Gamma} A^\Gamma_{lmn}\right) {}^*\!\bar{ \mathcal{F}}_{4\Sigma}\right]+\text{c.c.}\right\}
\\
&-\frac{1}{3!}\partial_k \left(\varepsilon^{klmn} \mathcal{N}_{\Lambda\Sigma}\, B^\Lambda_{lmn} {}^*\! G_4^\Sigma\right) \, .
\end{split}
\end{equation}

We notice that in \eqref{N2LDD} no gauging parameter appears. Indeed, the entire superpotential is dynamically generated once the gauge three-forms are set on-shell. In fact, from \eqref{N2NLScomm}, the equations of motion for the real gauge three-forms $A^\Lambda_3$ and $\tilde A_{3\,\Lambda}$ give
\begin{equation}
\begin{split}
\text{Re}\, \left(\mathcal{N}^{\Lambda\Sigma}  {}^*\!\bar{\mathcal{F}}_{4\,\Sigma}\right) =- m^\Lambda\,,\qquad \text{Re}\, \left(\mathcal{N}^{\Sigma\Gamma}  \bar F_{\Lambda\Sigma} {}^*\!\bar{ \mathcal{F}}_{4\,\Gamma}\right) = e_\Lambda\,,
\end{split}
\end{equation}
with $e_\Lambda$ and $m^\Lambda$ arbitrary real constants. These equations can be recast as
\begin{equation}
\label{N2Nsolcalff}
{}^*\! \mathcal{F}_{4\,\Lambda} = - i(e_\Lambda+ \bar F_{\Lambda\Sigma} m^\Sigma)\,.
\end{equation}
The equation of motion for the three-forms $B_3^\Lambda$ gives
\begin{equation}
\label{N2Nsolhh}
{}^*\! G^\Lambda_4 = \frac12 \mathcal{N}^{\Lambda\Sigma} \xi_\Sigma
\end{equation}
with $\xi_\Sigma$ arbitrary real constants.
Plugging \eqref{N2Nsolcalff} and \eqref{N2Nsolhh} in \eqref{N2NLScomm} we obtain directly
the same potential as \eqref{VpotN=2}, for the particular choice \eqref{N2LagGauging}.

\section{Dynamical Partial Supersymmetry Breaking}
\label{sec:DPSB}

In this section we examine how the partial breaking of supersymmetry is realized in terms of the gauge three-forms. We discuss to original model of \cite{Antoniadis:1995vb} using our formalism and then we construct an effective theory capturing its low energy description, along the lines of \cite{Bagger:1996wp,Rocek:1997hi, Antoniadis:2017jsk}. Through the entire discussion, all the parameters are going to be generated dynamically.

\subsection{Dynamical Antoniadis--Partouche--Taylor model} 
We consider the Lagrangian \eqref{N2LagSS} which, as we have shown, reproduces the scalar potential \eqref{N2LV}. This is the same scalar potential of \cite{Antoniadis:1995vb}, therefore we can argue that \eqref{N2LagSS} is a different off-shell completion of the same on-shell model. In \cite{Antoniadis:1995vb} it is discussed how the scalar potential \eqref{N2LV} admits vacua in which the $\mathcal{N}=2$ supersymmetry is partially spontaneously broken and a massless goldstino is present in the spectrum, together with a massive scalar and a massive fermion. We refer therefore the reader to \cite{Antoniadis:1995vb,Antoniadis:2017jsk} for additional details concerning the nature of the vacua of \eqref{N2LV}, while in the present subsection we concentrate on the analysis of the supersymmetry transformations of the fermions, in our formalism, in order to identify the goldstino. With respect to \cite{Antoniadis:1995vb}, we are going to give conditions on the gauge three-forms which are valid off-shell and which match the result of \cite{Antoniadis:1995vb}, when going on-shell.

To tell what amount of supersymmetry is preserved, it is necessary to examine how the supersymmetry variations of the fermions behave. We then focus on
\begin{equation}
\label{SusyVar2PsiPB}
\delta {\bm \Psi} =\frac{i}{\sqrt{2}} (\vec{\sigma} \cdot \vec{Y}) {\bm \eta} + \ldots\,.
\end{equation}
and we remind that a goldstino transforms with a shift under the broken supersymmetry.
It is clear that, for generic values of $\vec{Y}$, the whole $\mathcal{N}=2$ supersymmetry is non-linearly realized and the vacuum is not supersymmetric.
In order for the vacuum to preserve $\mathcal{N}=1$ supersymmetry, it is therefore necessary that a linear combination of the fermions $\psi$ and $\lambda$ transforms homogeneously, namely without a shift in any of the two supersymmetry parameters.
This can happen if and only if the matrix
\begin{equation}
\label{SusyVar2Matr}
\vec{\sigma} \cdot \vec{Y}  = \begin{pmatrix} Y^3 & Y^1-i Y^2 \\Y^1 + i Y^2 & -Y^3\end{pmatrix}
\end{equation}
has at least one zero eigenvalue. A necessary condition is that its determinant
\begin{equation}
\label{detsy}
{\rm det}\, ( \vec{\sigma} \cdot \vec{Y} ) = - \vec{Y} \cdot \vec{Y}
\end{equation}
vanishes. 
As pointed out in \cite{Antoniadis:1995vb}, when the magnetic Fayet--Iliopoulos parameter is turned off in \eqref{N2Lag}, then $\vec{Y}$ is real and the quantity \eqref{detsy} is always positive, but for the trivial case in which $\vec{Y}=0$ and supersymmetry is totally preserved. When the magnetic gauging are inserted, however, 
$\vec{Y}$ acquires a non-vanishing imaginary part and the matrix $ \vec{\sigma} \cdot \vec{Y} $ can be degenerate. In this case, the vacuum can preserve  $\mathcal{N}=1$ supersymmetry. This is how the partial breaking $\mathcal{N}=2 \rightarrow \mathcal{N}=1$ was originally conceived in \cite{Antoniadis:1995vb}, but here we adopt a slightly different perspective, in which the mechanism is sourced by a certain choice of boundary conditions, compatible with the equations of motions.

Considering \eqref{N2LagSS}, the auxiliary fields $\vec{Y}$ obtained form the variant multiplets \eqref{Alt_S} depend on the gauge three-forms as
\begin{equation}
\label{Y3forms}
\vec{Y} =-2 \begin{pmatrix}
{\rm Re}\, {}^*\! F_4 + m\\ -{\rm Im}\,{}^*\! F_4 +i m\\ \frac{1}{\sqrt{2}} {}^*\! G_4  \end{pmatrix}\,.
\end{equation}
Therefore, it is possible to realise that the matrix
\begin{equation}
\label{SusyVar2Matr4f}
\vec{\sigma} \cdot \vec{Y}  = -2 \begin{pmatrix} \frac{1}{\sqrt{2}} {}^*\! G_4  & {}^*\! F_4 + 2m\\{}^*\! \bar{F}_4 & -\frac{1}{\sqrt{2}} {}^*\! G_4 \end{pmatrix}
\end{equation}
is degenerate when
\begin{equation}
\frac12 ({}^*\! G_4)^2 = -|{}^*\! F_4|^2 - 2m\, {}^*\! \bar{F}_4  \,.
\end{equation}
This equation can be solved by
\begin{equation}
\label{DegenF4}
{\rm Im}\,{}^*\! {F}_4 =0 \qquad {\rm and} \qquad \frac12 ({}^*\! G_4)^2 = -2m\,{\rm Re}\,{}^*\! {F}_4  -  |{}^*\! F_4|^2\,,
\end{equation}
which is are off-shell conditions on the gauge three-forms for the partial breaking to occur. When going on-shell by using \eqref{Alt_Fos} and \eqref{Alt_Gos}, $\vec{Y}$ becomes
\begin{equation}
\label{Y3formsOS}
\vec{Y} = \frac{2}{\text{Im}\, F_{SS}}\begin{pmatrix}
0\\ e+ m \bar F_{SS} \\ -\frac{\xi}{2\sqrt{2}} \end{pmatrix}
\end{equation}
and, in order for \eqref{DegenF4} to be satisfied, we need
\begin{equation}
\label{PB_Vacuum}
\frac{e}{m} = - \text{Re}\, F_{SS}\,,\quad  \quad  -\frac{\xi}{2\sqrt{2}\, m} =  \text{Im}\, F_{SS}
\end{equation}
so that \eqref{Y3formsOS} gives
\begin{equation}
\vec{Y} = 2m\begin{pmatrix}
0\\ -i \\  1
\end{pmatrix},
\end{equation}
which matches the result in \cite{Antoniadis:1995vb}.
For this choice of $\vec{Y}$, the matrix \eqref{SusyVar2Matr} is degenerate. 
The on-shell supersymmetry transformations are indeed
\begin{equation}
\begin{aligned}
\delta \psi = i m (-\eta_1+\eta_2)+\ldots\,,\quad 
\delta \lambda = i m (\eta_1-\eta_2)+\ldots
\end{aligned}
\end{equation}
and we recognize that the combination $\psi+\lambda$ transforms linearly
\begin{equation}
\delta (\psi+\lambda) = 0 +\ldots,
\end{equation}
while the combination $\psi-\lambda$ transforms non-linearly
\begin{equation}
\delta (\psi-\lambda) = -2im (\eta_1-\eta_2) +\ldots,
\end{equation}
signaling the presence of a goldstino and the spontaneous breaking of one of the two supersymmetries. In our formalism, therefore, the partial breaking of supersymmetry is a consequence of the boundary conditions \eqref{Alt_Fos} and \eqref{Alt_Gos}. A different choice of the boundary conditions would generically lead to a different amount of broken supersymmetry.

\subsection{The low energy effective description}

In this subsection we use the formalism we have developed in order to construct an effective description for the previous model with partially broken supersymmetry, along the lines of \cite{Bagger:1996wp, Rocek:1997hi, Antoniadis:2017jsk, Dudas:2017sbi}. We will give evidence that the boundary terms that we have been including so far in all the Lagrangians are not solely an artifact of the formalism, but they contain important physical information.

The starting point is the Lagrangian \eqref{Alt_N2LD}, or equivalently \eqref{N2LagSS}. For convenience, however, this time we choose the boundary conditions so as the solutions to the equations of motion \eqref{Alt_Fos} and \eqref{Alt_Gos} now read
\begin{align}
\label{Alt_Fos2}
F_{klmn} &=-\frac{i}{{\rm Im}\, F_{SS}}\left(e+i \frac{\xi}{2\sqrt 2}+m F_{SS}\right)\varepsilon_{klmn}\,,
\\
\label{Alt_Gos2}
G_{klmn} &= 0\,.
\end{align}
The Lagrangian associated to these new boundary conditions and the one studied in the previous sections are related by a $\rm SU(2)_R$ transformation and therefore their physical properties are not changed.\footnote{In terms of the standard formalism for treating $\mathcal{N}=2$ Lagrangians, with respect to the previous $\rm SU(2)$ gauge choice, the Fayet--Iliopoulos parameter $\xi$ has become now the imaginary part of the electric gauging parameter $e$. 
} The scalar potential is given by
\begin{equation}
\mathcal{V} = \frac{1}{\text{Im}\, F_{SS}}\left| e + i\frac{\xi}{2\sqrt 2}+m F_{SS}\right|^2
\end{equation}
and differs from \eqref{N2LV} only for an irrelevant additive constant.
In particular the conditions \eqref{DegenF4} and \eqref{PB_Vacuum} for the existence of a vacuum with partial breaking of supersymmetry are again satisfied, but the triplet of auxiliary fields $\vec Y$ is now rotated to
\begin{equation}
\vec{Y} = \frac{2}{\text{Im}\, F_{SS}}\begin{pmatrix}
\frac{\xi}{2\sqrt{2}}\\ e+ m \bar F_{SS} \\ 0 \end{pmatrix}\,,
\end{equation}
which, setting the three-forms on-shell, becomes
\begin{equation}
\vec{Y} = 2m\begin{pmatrix}
-1\\ -i \\  0
\end{pmatrix}.
\end{equation}
The spectrum in this vacuum is the same as before: it contains a massless goldstino, a massive scalar and a massive
fermion, whose masses are proportional to $F_{SSS}$.

It is possible then to construct an effective theory for this setup by restricting the analysis to an energy regime well below the scale given by the masses of the massive fields, or equivalently by taking the formal limit $F_{SSS}\to \infty$.
To this purpose, we first expand the Lagrangian \eqref{Alt_N2LD} around the vacuum. In particular we assume that the choices 
\begin{subequations}\label{BG_Vac}
\begin{align}
F_{SS}^{(0)} &= - \frac{e}{m} - i\frac{\xi}{2 \sqrt{2} m}\,, \label{BG_VacA}
\\
F_{klmn}^{(0)} &=-\frac{i}{{\rm Im}\, F_{SS}^{(0)}}\left(e+i\frac{\xi}{2\sqrt 2}+m F_{SS}\right)\varepsilon_{klmn}\,,\label{BG_VacB}
\\
G_{klmn}^{(0)} &=  0\label{BG_VacC}
\end{align}
\end{subequations}
hold for a particular background value $S_0$ of the superfield $S$, set $S= S_0 + \tilde S$ and expand around $S_0$. 
Expanding then \eqref{Alt_N2LD}, along with the boundary terms \eqref{Alt_N2LDbd}, using \eqref{BG_Vac}, we arrive at the following effective Lagrangian
\begin{equation}
\label{Lfluct}
\begin{aligned}
\mathcal{L} &= \left\{F_{SS}^{(0)}  \int d^2 \theta\, \left(  \frac{i}{8} \tilde{S} \bar D^2 \bar{\tilde{S}} - \frac{i}4  W^{\alpha} W_{\alpha}+ m \tilde S \right)+\left(e+ i \frac{\xi}{2\sqrt{2}}\right)\int d^2 \theta\, \tilde{S}+  \text{c.c.} \right\}\\
&\quad\,+\mathcal{L}_{\rm bd} +\ldots\,,
\end{aligned}
\end{equation} 
with
\begin{equation}
\begin{split}
\mathcal{L}_{\rm bd}= &\frac14 \left(\int d^2\theta \bar D^2 - \int d^2\bar\theta D^2\right) \left[ {\rm Im}\, F_{SS}^{(0)} \left(\frac 14 {D}^2 {\tilde S}   \right) {\tilde\Sigma}_1\right]+\text{c.c.} \, ,
\end{split}
\end{equation}
where the dots stand for higher order terms in the fluctuations.
We notice that the coupling linear in $\tilde S$ with the complex electric parameter $e + i\frac{\xi}{2\sqrt 2}$ is generated from the boundary terms \eqref{Alt_N2LDbd}, after we set
\begin{equation}
\Lambda_1 = \Lambda_1^{(0)} + \tilde \Lambda_1 = \left(e + i\frac{\xi}{2\sqrt 2}\right)+ \tilde \Lambda_1\,.
\end{equation}
The presence of this coupling is crucial for the existence of the effective theory and it is indeed a pure boundary term contribution.

We take now the limit of infinite mass. By setting to zero the divergent part of the equations of motion of the fluctuations, the following constraint is produced 
\begin{equation}
\label{constr1}
\frac i8 S \bar D^2 \bar S - \frac i4 W^\alpha W_\alpha + m S =0
\end{equation}
where from now on we omit the tilde on the fields.
This constraint can be recast into the form
\begin{equation}
S = \frac{W^\alpha W_\alpha}{-4im + \frac12 \bar D^2 \bar S},
\end{equation}
which can be solved iteratively \cite{Bagger:1996wp, Rocek:1997hi}. The solution is
\begin{equation}
S =\frac{i}{4m} \left\{W^2 -  \bar D^2\left[\frac{W^2 \bar W^2}{(4m)^2+ A + \sqrt{(4m)^4+2 (4m)^2A+ B^2}}\right]\right\},
\end{equation}
where we have defined
\begin{align}
A  =\frac{D^2 W^2 + \bar D^2 \bar W^2}{2}\,, \qquad B  =\frac{D^2 W^2 - \bar D^2 \bar W^2}{2}\,.
\end{align}
It is known \cite{Rocek:1997hi,Ferrara:2014oka, Antoniadis:2017jsk, Dudas:2017sbi} that, in the language of $\mathcal{N}=2$ superspace, the constraint \eqref{constr1} is a consequence of a nilpotent constraint imposed on top of the original vector superfield, namely $\mathcal{S}^2=0$. This constraint is removing the entire $\mathcal{N}=1$ chiral superfield $S$ from $\mathcal{S}$ and expresses it as a function of the remaining $\mathcal{N}=1$ vector superfield $W_\alpha$.

By implementing the constraint into \eqref{Lfluct}, the low energy effective action reduces
\begin{equation}
\begin{aligned}
\label{LeffBG}
\mathcal{L} &= \left( \frac{i e}{4m} - \frac{\xi}{8 \sqrt{2} m}\right)\int d^2 \theta\,\left\{W^2 -  \bar D^2\left[\frac{W^2 \bar W^2}{(4m)^2+ A + \sqrt{(4m)^4+2 (4m)^2A+ B^2}}\right]\right\}\\
&\quad\, + \text{c.c.}\,.
\end{aligned}
\end{equation}
We stress that, from our perspective, this action is entirely contained in the boundary terms \eqref{Alt_N2LDbd}, which captures therefore the effective description of the model in the infrared regime. As a consequence, when considering four-dimensional Lagrangian with gauge three-forms, it is important to include the appropriate boundary terms, since they can contain non-trivial physical information. 
The on-shell bosonic components of \eqref{LeffBG} can be recast into the form of a Born--Infeld action
\begin{equation}
\label{LDBI}
\mathcal{L} = -\frac{m\xi }{\sqrt 2}\left(1-\sqrt{-\det \left(\eta_{mn}+F_{mn}\right)}\right)+\frac{me}{4}\epsilon^{mnpq}F_{mn}F_{pq},
\end{equation}
where we have rescaled $F_{mn} \to \sqrt 2 m F_{mn}$ in order to have canonically normalized kinetic terms. We stress that all the parameters appearing in this Lagrangian have been generated dynamically, as vacuum expectation values of gauge three-forms. In particular, the product $m \xi$ may be related to the tension of the D3 brane described by \eqref{LDBI}, for which we have provided a dynamical origin.
We notice finally that, by choosing a boundary condition in which $m=0$, the partial breaking of supersymmetry does not occur and this effective description does not hold.

\section{Conclusion}

In this work we have studied $\mathcal{N}=2$ global supersymmetric models in which the parameters entering the superpotential and the scalar potential have a dynamical origin. A systematic procedure has been given in order to trade standard $\mathcal{N}=2$ multiplets for variant versions, in which gauge-three forms appear as non-propagating degrees of freedom. When going on-shell, parameters are generated in the theory as integration constants for the gauge three-forms. In other words, the choice of parameters in the original Lagrangian is traded for the problem of specifying certain boundary conditions for the gauge three-forms, compatibly with their equations of motion.

Our results may be relevant, first of all, for understanding the origin of parameters in effective theories which come from string theory.  It is known in fact that string theory does not have any free parameter, but the string length and therefore eventual additional parameters appearing in models originating from string theory have to be interpreted as expectation values of certain fields.
In this context, four dimensional theories preserving $\mathcal{N}=2$ supersymmetry may appear for example when compactifying string theory on Calabi--Yau three folds.

The results presented in this work can be of interest also for the study of supersymmetric theories from a pure four-dimensional point of view, in particular for understanding the relation between off-shell and on-shell formulations of extended supersymmetry, which has not been completely clarified at present. We have presented in fact novel off-shell multiplets and Lagrangians which reproduce correctly known on-shell setups.

As possible further directions, it would be important to extend the analysis to local supersymmetry and explore the coupling of the gauge three-forms to membranes already in four dimensions as in \cite{Bandos:2018gjp}. We leave these developments for future work.


\subsection*{Acknowledgements}

N.C. is supported by an FWF grant with the number
P 30265. S.L. is grateful for hospitality to Istituto de F\'isica Te\'orica UAM-CSIC, Madrid, where this work was completed.
We thank Fotis Farakos, Dmitri Sorokin, Gabriele Tartaglino-Mazzucchelli and especially Gianguido Dall'Agata and Luca Martucci for discussions and comments on the manuscript.


\appendix

\section{Conventions}
\label{app:Conv}
The components of a three-form $A_3$ are defined as
\begin{equation}
A_3 = \frac{1}{3!} A_{kmn} d x^k\, \wedge d x^m\, \wedge d x^n\,,
\end{equation}
whose field strength is defined as
\begin{equation}
F_4 \equiv d A_3\,, \qquad F_4 = \frac{1}{4!} F_{klmn}  d x^k\, \wedge d x^l\, \wedge d x^m\,\wedge d x^n
\end{equation}
with components
\begin{equation}
F_{klmn} = 4\, \partial_{[k} A_{lmn]}\,.
\end{equation}
The Hodge-dual of a four-form field strength $F_4$ is
\begin{equation}
{}^*\!F_4 = \frac{1}{4!}\varepsilon^{klmn} F_{klmn} = \frac{1}{3!} \varepsilon^{klmn}  \partial_{[k} A_{lmn ]}\,
\end{equation}
and in our conventions
\begin{equation}
\varepsilon_{klmn} \varepsilon^{pqrs} = - 4! \delta^{p}_{[k} \delta^{q}_{l} \delta^{r}_{m}\delta^{s}_{n]}\,.
\end{equation}

\section{Component structure of ${\cal N}=1$ superfields}
\label{app:Super}

Here we collect the component structures of the $\mathcal{N}=1$ multiplets considered throughout this work.

The chiral multiplet $X$ is defined by 
\begin{equation}
\bar{D}_{\dot\alpha} X = 0
\label{eq:ChiralA}
\end{equation}
and its component expansion is 
\begin{equation}
X = \varphi + \sqrt{2} \theta \psi + \theta^2 f + i \theta \sigma^m \bar\theta \partial_m \varphi -\frac{i}{\sqrt{2}} \theta^2 \partial_m \psi \sigma^m \bar\theta + \frac 14 \theta^2\bar\theta^2 \Box \varphi,
\label{ChiralB}
\end{equation}
where $\varphi$ and $f$ are complex scalar fields, while $\psi$ is a Weyl spinor. 
The independent components of $\Phi$ can be defined by the projections
\begin{equation}\label{ChiralComp}
\begin{aligned}
\Phi | &= \varphi \,, \\
D_\alpha \Phi &|= \sqrt 2 \psi_\alpha \,,\\
-\frac14 D^2 \Phi| &= f \,,
\end{aligned}
\end{equation}
where the vertical line means that the quantity is evaluated at $\theta=\bar\theta=0$.
	
The real scalar multiplet $V$ is defined by $V = \bar V$ and it has the following component structure 
\begin{equation}
\begin{split}
V =& \, u + i \theta \chi  - i \bar\theta \bar\chi + {i} \theta^2 \bar\varphi -  {i} \bar\theta^2 {\varphi} - \theta \sigma^m \bar\theta v_m  \\
& +i\theta^2\bar{\theta} \left( \bar{\lambda} 
+\frac{i}{2}\bar{\sigma}^m\partial_m \chi\right)-i\bar{\theta}^2 \theta \left( \lambda+\frac{i}{2}{\sigma}^m\partial_m \bar{\chi}\right)+\frac12\theta^2\bar\theta^2 \left({\rm D}-\frac12\Box u \right) \, , 
\label{VectorB}
\end{split}
\end{equation}
where $u$ and $D$ are real scalar fields, $\varphi$ is a complex scalar field, $v_m$ is a real vector field and $\chi$ and $\lambda$ are Weyl spinors. 
The independent components of $V$ can be defined by projections
\begin{equation}
\label{Vprojections}
\begin{aligned}
V | &= u \,,\\
D_\alpha V| &= i \chi_\alpha\,,\\
\frac 14 \bar{\sigma}^{\dot\alpha \alpha}_m [D_\alpha, \bar{D}_{\dot\alpha}] V| &= v_m\,,\\ 
\frac{i}{4} D^2 V| &= \bar\varphi\,,\\
-\frac14 \bar D^2 D_\alpha V| &= -i\lambda_\alpha\,,\\
\frac{1}{16} D^2 \bar{D}^2 V| &= \frac12\left({\rm D} - i \partial^m v_m \right) \, .
\end{aligned}
\end{equation}

The real linear multiplet $L$ is a real multiplet which, in addition, satisfies the condition
\begin{equation}
D^2 L = 0\,, \qquad \bar{D}^2 L = 0 \,.
\label{RealLMdef}
\end{equation}
Its component expansion is
\begin{equation}
\begin{split}
L = &l + i \theta \eta -i \bar\theta \bar\eta - \frac12 \theta \sigma_m \bar\theta \varepsilon^{mnpq} \partial_{n}\Lambda_{pq} \\
&+ \frac12 \theta^2\bar\theta \bar\sigma^m \partial_m \eta - \frac12  \bar\theta^2  \theta \sigma^m \partial_m \bar\eta -\frac14 \theta^2\bar\theta^2 \Box l \, , 
\end{split}
\label{RealLM}
\end{equation}
where $l$ is a real scalar,  $\Lambda_{mn}$ is a rank 2 antisimmetric tensor  and $\eta$ is a Weyl spinor. \\
The independent components of $L$ can be defined by projections 
\begin{equation}
\begin{aligned}
L | &= l \,, \\
D_\alpha L| &= i \eta_\alpha\,,\\
\frac12\bar{\sigma}^{m\,\dot\alpha\,, \alpha}\left[D_\alpha,\bar{D}_{\dot{\alpha}}\right] L| &= \varepsilon^{mnpq} \partial_{n}\Lambda_{pq} \,.
\end{aligned}
\end{equation}

The complex linear multiplet $\Sigma$ satisfies the condition
\begin{equation}
\bar{D}^2 \Sigma = 0 \,.
\label{ComplexLMdef}
\end{equation}
Its component expansion is 
\begin{equation}
\begin{split}
\Sigma = &\sigma + \sqrt 2\theta \psi + \sqrt{2} \bar\theta \bar\rho - \theta \sigma_m \bar\theta \mathcal{B}^{m} + \theta^2 \bar s+ \sqrt 2\theta^2\bar\theta \bar\zeta \\
&-\frac{i}{\sqrt{2}} \bar\theta^2 \theta \sigma^m \partial_m \bar\rho + \theta^2\bar\theta^2 \left(\frac{i}{2} \partial_m \mathcal{B}^{m} -\frac14 \Box \sigma \right).
\end{split},
\label{eq:ComplexLM}
\end{equation}
where $\sigma$ and $\bar s$ are complex scalars, $\rho$, $\psi$ and $\xi$ are Weyl spinors and $\mathcal{B}^{m}$ is a complex vector which is Hodge dual to the three-form 
\begin{equation}
\mathcal{B}^{m} = \frac{1}{3!} \varepsilon^{mnpq} \mathcal{B}_{npq}.
\label{ComplexLMvector}
\end{equation}
The components of $\Sigma$ can be defined  by the projections
\begin{equation}
\label{SigmaCom}
\begin{aligned}
\Sigma | &= \sigma \,,\\
D_\alpha \Sigma| &= \sqrt 2 \psi_\alpha\,,\\
\bar D_{\dot \alpha} \Sigma| &= \sqrt 2 \bar \rho_{\dot\alpha}\,,\\
\frac14 \bar{\sigma}^{m\,\dot\alpha \alpha}
\left[D_\alpha,\bar{D}_{\dot{\alpha}}\right] \Sigma| &= \mathcal{B}^{m}\,,\\
-\frac14 D^2 \Sigma| &= \bar s\,,\\
\bar D_{\dot \alpha} D^2 \Sigma| & = -4\sqrt 2 \bar \zeta_{\dot \alpha} +2\sqrt 2 i \, \partial_m \psi^\alpha \sigma^m_{\alpha \dot \alpha}\,,\\
\frac{1}{16} \bar{D}^2 D^2 \Sigma| &= i \partial_m \mathcal{B}^m \,.
\end{aligned}
\end{equation}

\section{Case with $N$ vector multiplets}
\label{app:Nvec}

The procedure outlined in Section \ref{sec:N23form1} can be generalized to the case involving an arbitrary number $N$ of vector multiplets. We recall that the $\mathcal{N}=2$ Lagrangian equipped with both electric and magnetic gauging parameters is
\begin{equation}
\tag{\ref{N2NLag}}
\mathcal{L} = \frac{i}{2} \int d^2 \theta d^2 \tilde \theta \left[F( \mathcal{A}) - \mathcal{A}_{D\,\Lambda} \mathcal{A}^\Lambda \right] + \frac12 \left(\vec{E}_\Lambda\cdot \vec{Y}^\Lambda+  \vec{M}^\Lambda\cdot \vec{Y}_{D \Lambda}\right)+\text{c.c.}\,,
\end{equation}
which, in $\mathcal{N}=1$ language, translates into
\begin{equation}
\label{AltN_LN=1a}
\begin{aligned}
\mathcal{L} &= -\frac i4 \int d^2\theta \left( F_{\Lambda \Sigma}  W^{\Lambda\,\alpha} W_\alpha^\Sigma - 2 W_{D\,\Lambda}^\alpha W_\alpha^\Lambda \right)\\
&\quad - \frac i2 \int d^4 \theta F_\Lambda \bar{X}^\Lambda + \frac 14 \int d^2 \theta \Phi^\Lambda \left( X_{D\,\Lambda} - F_\Lambda\right) +\frac12 \left(\vec{E}_\Lambda\cdot \vec{Y}^\Lambda+  \vec{M}^\Lambda\cdot \vec{Y}_{D\, \Lambda}\right)+ {\rm c.c.}\,.
\end{aligned}
\end{equation}
We use the $\rm SU(2)$ R-symmetry of the theory to rotate the parameters such that
\begin{equation}
\label{N2NLagGauging}
\Re \vec{E}_\Lambda =\left(0,-e_\Lambda,\frac{\xi_\Lambda}{2\sqrt{2}}\right)\,, \qquad \vec{M}^\Lambda = (0,-m^\Lambda,0)\,,
\end{equation}
with $e_\Lambda$, $m^\Lambda$ and $\xi_\Lambda$ real constants. In this way the Lagrangian can be expressed in the $\mathcal{N}=1$ Language as
\begin{equation}
\begin{aligned}
\label{N2NLagB}
\mathcal{L} = &\int d^4\theta\, K(X,\bar X)+ \left(\frac14 \int d^2\theta\, \tau_{\Lambda \Sigma} (X) W^{\Lambda\alpha} W^\Sigma_\alpha+\text{c.c.}\right)\\
&+\left(\int d^2\theta\,W(X)+\text{c.c.}\right) + \xi_\Lambda \int d^4\theta\,V^\Lambda \,,
\end{aligned}
\end{equation}
with 
\begin{equation}
\tau_{\Lambda \Sigma}(X) =- i F_{\Lambda \Sigma}, \qquad K = \frac{i}{2}\left(X^\Lambda \bar F_{ \Lambda} - \bar X^{\Lambda} F_\Lambda \right),
\end{equation}
and the superpotential $W(X)$ is given by
\begin{equation}
\label{N2NW}
W(X)= e_\Lambda X^\Lambda + m^\Lambda F_\Lambda(X)\,.
\end{equation}
The bosonic components are
\begin{equation}
\label{N2NLScom}
\begin{split}
\mathcal{L}\big|_{\text{bos}} =& -\mathcal{N}_{\Lambda \Sigma}\, \partial_m \varphi^\Lambda\, \partial^m \bar \varphi^\Sigma -\frac 14 \mathcal{N}_{\Lambda \Sigma} F^{\Lambda\, mn}F^\Sigma_{mn}  - \frac18 \text{Re}\, F_{\Lambda \Sigma}\, \varepsilon_{klmn} F^{\Lambda\,kl} F^{\Sigma mn} \\
&+  \mathcal{N}_{\Lambda \Sigma} f^\Lambda \bar f^\Sigma    +\frac12  \mathcal{N}_{\Lambda \Sigma} D^\Lambda D^\Sigma\\
& + (e_\Lambda +  F_{\Lambda \Sigma} m^\Sigma) f^\Lambda + (e_\Lambda +\bar F_{\Lambda \Sigma} m^\Sigma) \bar f^\Lambda + \frac{1}{2} \xi_\Lambda D^\Sigma\,.
\end{split}
\end{equation}
Integrating out the auxiliary fields $f^\Lambda$ and $D^\Lambda$ we arrive at
\begin{equation}
\label{N2NLScomOS}
\begin{aligned}
\mathcal{L}\big|_{\text{bos}} =& -\mathcal{N}_{\Lambda \Sigma}\, \partial_m \varphi^\Lambda\, \partial^m \bar \varphi^\Sigma -\frac 14 \mathcal{N}_{\Lambda\Sigma} F^{\Lambda\,mn}F^\Sigma_{mn} \\
& - \frac18 \text{Re}\, F_{\Lambda \Sigma}\, \varepsilon_{klmn} F^{\Lambda\, kl} F^{\Sigma\, mn} - \mathcal{V}(\varphi, \bar \varphi)\,,	
\end{aligned}
\end{equation}
with the scalar potential
\begin{equation}
\label{N2NV}
\begin{split}
\mathcal{V}(\varphi, \bar \varphi) =\; &\mathcal{N}_{\Lambda \Sigma} m^\Lambda m^\Sigma + \mathcal{N}^{\Lambda \Sigma} (e_\Lambda + m^\Gamma \text{Re}\,F_{\Gamma\Lambda})(e_\Sigma + m^\Delta \text{Re}\, F_{\Sigma \Delta})+\frac18 \mathcal{N}^{\Lambda \Sigma} \xi_\Lambda \xi_\Sigma\,,	
\end{split}
\end{equation}
which coincides with \eqref{VpotN=2} for the particular choice \eqref{N2NLagGauging}.	

The first step of the procedure consists in promoting the magnetic gauging parameters $\vec{M}^\Lambda$ to be dynamical. This may be achieved by trading the Lagrange multiplier vector multiplets $\mathcal{A}_{D\,\Lambda}$ for the variant three-form multiplets \eqref{ChiralS}. At $\mathcal{N}=1$ level, we promote then the term $\vec{M}^\Lambda\cdot \vec{Y}_{D\, \Lambda}$ in \eqref{AltN_LN=1a} to a full dynamical entity as
\begin{equation}
\label{AltN_LN=1D}
\begin{aligned}
\mathcal{L} &= \bigg\{-\frac i4 \int d^2\theta \left( F_{\Lambda\Sigma}  W^{\Lambda\,\alpha} W_\alpha^\Sigma - 2 W_{D\,\Lambda}^\alpha W_\alpha^\Lambda \right)-
\\
& \quad\quad- \frac i2 \int d^4 \theta F_\Lambda \bar{X}^\Lambda + \frac 14 \int d^2 \theta \Phi^\Lambda \left( X_{D\,\Lambda} - F_\Lambda\right) + \frac12 \vec{E}_\Lambda \cdot \vec{Y}^\Lambda+ {\rm c.c.}\bigg\}+
\\
&\quad+\left\{\int d^2 \theta\, \left(\Lambda_1^{D\,\Pi} X_{D\,\Pi} + \frac14   \bar D^2 (\Sigma_{1 {D\,\Pi}} \bar \Lambda_1^{D\,\Pi}) \right)+ {\rm c.c.}\right\}
\\
&\quad+\left\{\frac18 \int d^2\theta\,\bar D^2 [\Lambda_2^{D\,\Pi} (\Sigma_{2{D\,\Pi}} - V_{D\,\Pi})] + \text{c.c.} \right\}.
\end{aligned}
\end{equation}
The third line provides the trading between the $\mathcal{N}=1$ chiral multiplets $X_{D\,\Pi}$ and a double three-form multiplet, while the fourth line the one between the $\mathcal{N}=1$ vector multiplets $W_{D\,\Pi}^\alpha$ and their three-form counterparts.
Integrating the complex linear superfields $\Sigma_{1{D\,\Pi}}$ and $\Sigma_{2{D\,\Pi}}$ gives
\begin{equation}
\label{AltN_LambdaOS}
\Lambda_1^{D\,\Pi} = -M^{2\,\Pi}-i M^{1\,\Pi}\,,\qquad \Lambda^{D\,\Pi}_2= 2\sqrt{2} M^{3\,\Pi}\,,
\end{equation}
with $\vec{M}^{\Pi}$ arbitrary real constants, establishing the equivalence between \eqref{AltN_LN=1D} and \eqref{AltN_LN=1a}.
On the other hand, the variations with respect to the Lagrange multipliers $\Lambda_1^{D\,\Pi}$ and $\Lambda^{D\,\Pi}_2$ give the relations
\begin{align}
\delta \Lambda_1^{D\,\Pi}: \quad X_{D\,\Pi} &= -\frac14 \bar D^2 \bar\Sigma_{1D\,\Pi} \equiv S_{D\,\Pi}\,, \label{AltN_XDsol}
\\
\delta \Lambda_2^{D\,\Pi}: \quad  V_{D\,\Pi} &=  \frac{\Sigma_{2D\,\Pi} + \bar \Sigma_{2D\,\Pi}}{2} \equiv U_{D\,\Pi}\,, \label{AltN_VDsol}
\end{align}
and those with respect to the ordinary $\mathcal{N}=1$ superfields $X_{D\,\Pi}$ and $V_{D\,\Pi}$ result in
\begin{align}
\label{AltN_L1Dsol}
\delta X_{D\,\Pi}: \quad \Lambda^{D\,\Pi}_1 &= -\frac 14 \Phi^\Pi\,,
\\
\label{AltN_L2Dsol}
\delta V_{D\,\Pi}: \quad \Lambda^{D\,\Pi}_2 &= - {\rm Im} (D^\alpha W^\Pi_\alpha)\,.
\end{align}
Plugging (\ref{AltN_XDsol}-\ref{AltN_L2Dsol}) in \eqref{AltN_LN=1D}, we get
\begin{equation}
\label{AltN_LN=1Db}
\begin{aligned}
\mathcal{L} &=\bigg\{-\frac i4 \int d^2\theta \left( F_{\Lambda\Sigma}  W^{\Lambda\,\alpha} W_\alpha^\Sigma - 2 W_{D\,\Lambda}^\alpha W_\alpha^\Lambda \right)-
\\
& \quad\quad- \frac i2 \int d^4 \theta F_\Lambda \bar{X}^\Lambda + \frac 14 \int d^2 \theta \Phi^\Lambda \left( S_{D\,\Lambda} - F_\Lambda\right) + \frac12 \vec{E}_\Lambda \cdot \vec{Y}^\Lambda+ {\rm c.c.}\bigg\}+ \mathcal{L}_{\rm bd}^{(D)}
\end{aligned}
\end{equation}
with
\begin{equation}\label{AltN_LN=1Dbd}
\begin{split}
\mathcal{L}_{\rm bd}^{(D)}= &-\frac14 \left(\int d^2\theta \bar D^2 - \int d^2\bar\theta D^2\right) \left[\left(-\frac 14 \Phi^\Lambda  \right) \bar\Sigma_{1D\,\Lambda}\right]\\
&+\frac{1}{16} \left(\int d^2\theta \bar D^2 - \int d^2\bar\theta D^2\right)  \left[ - \text{Im} (D^\alpha W^\Lambda_\alpha) \Sigma_{2D\,\Lambda} \right] +\text{c.c.} \, .
\end{split}
\end{equation}

This Lagrangian can also be rewritten in $\mathcal{N}=2$ superspace as
\begin{equation}
\label{AltN_LN=2}
\mathcal{L} = \frac{i}{2} \int d^2 \theta d^2 \tilde \theta \left[F( \mathcal{A}) -  \mathcal{S}_{D\,\Lambda}  \mathcal{A}^\Lambda \right] +  \frac12 \vec{E}_{\Lambda}\cdot \vec{Y}^\Lambda + {\rm c.c.} + \mathcal{L}_{\rm bd}^{(D)}\,.
\end{equation}
where the $\mathcal{N}=2$ Lagrange multiplier $\mathcal{S}_{D\,\Lambda} $ is 
\begin{equation}\label{ChiralSDN}
\begin{split}
\mathcal{S}_{D\,\Lambda}(y,\theta, \tilde \theta) &= S_{D\,\Lambda} (y,\theta) + \sqrt 2 \tilde\theta^\alpha W_{{D\,\Lambda}\alpha}(y,\theta) + \frac14\tilde \theta^2  \bar D^2\bar S_{D\,\Lambda}\,.
\end{split}
\end{equation}
where $W_{{D\,\Lambda}\alpha} = -\frac14 \bar D^2 D_\alpha U_{D\,\Lambda}$, with $S_{D\,\Lambda}$ and $U_{D\,\Lambda}$ defined respectively in \eqref{AltN_XDsol} and  \eqref{AltN_VDsol}.

In order to re-express the Lagrangian \eqref{AltN_LN=2} only in terms of the chiral multiplets $\mathcal{A}^\Lambda$, the variant Lagrange multipliers $\mathcal{S}_{D\,\Lambda}$ have to be integrating out. This amounts in integrating out the complex linear superfields $\Sigma_{1 {D\,\Lambda}}$ and $\Sigma_{2 {D\,\Lambda}}$ which, as in \eqref{AltN_L1Dsol} and \eqref{AltN_L2Dsol}, results in setting
\begin{equation}
\label{AltN_DPhi_OS}
{\rm Im}\, {\rm D}^\Pi = \sqrt{2} M^{3\,\Pi}\, \qquad \Phi^\Pi = 4 (M^{2\,\Pi} +i M^{1\,\Pi})
\end{equation}
If we choose $\vec{M}^\Pi = (0,-m^\Pi,0)$, the Lagrangian \eqref{AltN_LN=1Db} reads
\begin{equation}
\label{AltN_LN=1c}
\begin{aligned}
\mathcal{L} &= 
-\frac i4 \int d^2\theta F_{\Lambda\Sigma}  W^{\Lambda\,\alpha} W^\Sigma_\alpha  - \frac i2 \int d^4 \theta F_\Lambda \bar{X}^\Lambda + m^\Lambda \int d^2 \theta  F_\Lambda  + \frac12 \vec{E}_\Lambda\cdot \vec{Y}^\Lambda+ {\rm c.c.}.
\end{aligned}
\end{equation}
In comparison with \eqref{AltN_LN=1D} and recalling the component structure of the chiral multiplet \eqref{N=2VecMult}, this suggests that integrating out $\mathcal{S}_{D\,\Lambda}$ constraints $\mathcal{A}^\Lambda$ to be the reduced chiral multiplet
\begin{equation}\label{AltN_A}
\mathcal{A}^\Lambda(y,\theta, \tilde \theta) = X^\Lambda(y,\theta) + \sqrt 2 \tilde\theta^\alpha W^\Lambda_\alpha(y,\theta) +\tilde \theta^2 \left(-2im^\Lambda+\frac14 \bar D^2\bar X^\Lambda\right).
\end{equation}

The second step is to generate dynamically also the electric gauging parameters $\vec{E}$, by promoting the reduced chiral multiplets $\mathcal{A}^\Lambda$ to three-form multiplets as well. We convert then \eqref{AltN_LN=1c} to
\begin{equation}\label{AltN_N2Master}
\begin{aligned}
\mathcal{L} =&  \int d^4\theta\, K(X,\bar X)+\left(\frac14 \int d^2\theta\, \tau_{\Lambda\Sigma}(X) W^{\Lambda\,\alpha} W^\Sigma_\alpha + \text{c.c.}\right)+\\
&+\left\{\int d^2\theta \Lambda_{1\Pi} X^\Pi
+\frac14 \int d^2\theta\, \bar D^2\left(\Sigma_{1\Pi} \bar \Lambda_1^\Pi\right)+ m^\Pi \int d^2 \theta F_\Pi+\text{c.c.}\right\}+\\
&+\left\{\frac18 \int d^2\theta\,\bar D^2 [\Lambda_{2\Pi} (\Sigma^\Pi_2 - V^\Pi)] + \text{c.c.} \right\}\,.
\end{aligned}
\end{equation}
As a check of consistency between the Lagrangians \eqref{AltN_N2Master} and \eqref{N2LagB}, we may integrate out the complex linear superfields $\Sigma_1$ and $\Sigma_2$, obtaining
\begin{equation}\label{AltN_Lambda_sol}
\Lambda_{1\Lambda} = e_\Lambda\,, \qquad \Lambda_{2\Lambda} = \xi_\Lambda\,,
\end{equation}
with $e_\Lambda$ and $\xi_\Lambda$ arbitrary real constants. Inserting \eqref{AltN_Lambda_sol} in \eqref{AltN_N2Master},  we in fact re-obtain \eqref{N2NLagB}.
Let us now recast the Lagrangian \eqref{AltN_N2Master} only in terms of the three-form multiplets. The variations with respect of the Lagrange multipliers $\Lambda_{1\Pi}$ and $\Lambda_{2\Pi}$ give
\begin{align}
\delta \Lambda_{1\Pi}: \quad X^\Pi &= -\frac14 \bar D^2 \bar\Sigma_1^\Pi \equiv S^\Pi\,, \label{AltN_Xsol}
\\
\delta \Lambda_{2\Pi}: \quad  V^\Pi &=  \frac{\Sigma_2^\Pi + \bar \Sigma_2^\Pi}{2} \equiv U^\Pi\,, \label{AltN_Vsol}
\end{align}
which trade the $\mathcal{N}=1$ multiplets for their three-form counterparts. The variations with respect to $X^\Pi$ and $V^\Pi$ give explicit expressions for the Lagrange multipliers $\Lambda_{1\Pi}$ and $\Lambda_{2\Pi}$:
\begin{align}
\label{AltN_L1sol}
\delta X^\Pi: \quad \Lambda_{1\Pi} &= \frac 14 \bar{D}^2 K_{\Pi} - \frac{1}{4} \tau_{\Sigma\Lambda\Pi} W^{\Sigma\,\alpha} W^\Lambda_\alpha- m^\Sigma F_{\Sigma\Pi}\,,
\\
\label{AltN_L2sol}
\delta V^\Pi: \quad \Lambda_{2\Pi} &= \mathcal{N}_{\Pi\Sigma} D^\alpha W^\Sigma_\alpha + \frac{1}{2} \left[ (D^\alpha \tau_{\Pi\Sigma}) W_\alpha^\Sigma - (\bar D_{\dot \alpha} \tau_{\Pi\Sigma}) \bar W^{\dot \alpha\, \Sigma} \right]\,.
\end{align}
Inserting (\ref{AltN_Xsol}--\ref{AltN_L2sol}) in \eqref{AltN_N2Master} we get
\begin{equation}
\label{AltN_N2LD}
\mathcal{L} = \int d^4\theta K(S,\bar S) + \left( \frac14 \int d^2\theta  \tau_{\Lambda\Sigma}(S) W^{\Lambda\,\alpha} W^\Sigma_\alpha + \text{c.c.}\right) +\mathcal{L}_{\rm bd},
\end{equation} 
with
\begin{equation}\label{AltN_N2LDbd}
\begin{split}
\mathcal{L}_{\rm bd}= &-\frac14 \left(\int d^2\theta \bar D^2 - \int d^2\bar\theta D^2\right) \left[\Lambda_{1 \Pi} \bar\Sigma_1^\Pi\right]\\
&+\frac{1}{16} \left(\int d^2\theta \bar D^2 - \int d^2\bar\theta D^2\right)  \left( \Lambda_{2\Pi}  \Sigma^\Pi_2 \right) +\text{c.c.} \, .
\end{split}
\end{equation}
with $\Lambda_{1\Pi}$ and $\Lambda_{2\Pi}$ specified by \eqref{AltN_L1sol} and \eqref{AltN_L2sol}.
The bosonic components of the Lagrangian \eqref{AltN_N2LD} are
\begin{equation}
\label{AltN_N2LScom}
\begin{split}
\mathcal{L}\big|_{\text{bos}} =&-\mathcal{N}_{\Lambda \Sigma}\, \partial_m \varphi^\Lambda\, \partial^m \bar \varphi^\Sigma -\frac 14 \mathcal{N}_{\Lambda \Sigma} F^{\Lambda\, mn}F^\Sigma_{mn}  - \frac18 \text{Re}\, F_{\Lambda \Sigma}\, \varepsilon_{klmn} F^{\Lambda\,kl} F^{\Sigma mn}   \\
&+ \mathcal{N}_{\Lambda \Sigma} {}^*\! {F}_4^\Lambda {}^*\!\bar{{F}}_4^\Sigma    +\frac12   \mathcal{N}_{\Lambda \Sigma}  {}^*\!G_4^\Lambda {}^*\!G_4^\Sigma+\left(-i m^\Lambda F_{\Lambda\Sigma} {}^*\!\bar{{F}}_4^\Sigma + {\rm c.c.}\right)+ \mathcal{L}_{\rm bd}
\end{split}
\end{equation}
with
\begin{equation}\label{AltN_N2Lbdcom}
\begin{split}
\mathcal{L}_{\rm bd} = &\left\{\frac{1}{3!} \partial_k\left[ B_{lmn}^\Lambda\left( \mathcal{N}_{\Lambda\Sigma}\bar F^{\Sigma\,klmn}- i m^\Sigma\, \varepsilon^{klmn} \bar F_{\Lambda\Sigma}\right)\right]+\text{c.c.}\right\}\\
&+\frac{1}{3!}\partial_k \left(\mathcal{N}_{\Lambda\Sigma}\, C_{lmn}^\Lambda  G^{\Sigma\,klmn}\right) \, .
\end{split}
\end{equation}
Setting the gauge three-forms on shell as
\begin{align}
\label{AltN_Fos}
F_{klmn}^\Lambda &=-i \mathcal{N}^{\Lambda\Sigma} (e_\Sigma+m^\Pi F_{\Pi\Sigma})\varepsilon_{klmn}\,,
\\
\label{AltN_Gos}
G_{klmn}^\Lambda &= -\frac12 \mathcal{N}^{\Lambda\Sigma}\, \xi_\Sigma\, \varepsilon_{klmn}\,,
\end{align}
the same potential as \eqref{N2NV} is recovered.

The Lagrangian \eqref{AltN_N2LD} can also be fully re-expressed in $\mathcal{N}=2$ language as
\begin{equation}
\label{N2LagNSS}
\mathcal{L} = \frac{i}{2} \int d^2 \theta d^2 \tilde \theta F( \mathcal{S})  +\text{c.c.} + \mathcal{L}_{\text{bd}}\,.
\end{equation}
provided that we introduce the $\mathcal{N}=2$ chiral multiplets defined as
\begin{equation}\label{AltN_S}
\mathcal{S}^\Lambda(y,\theta, \tilde \theta) = S^\Lambda(y,\theta) + \sqrt 2 \tilde\theta^\alpha W^\Lambda_\alpha(y,\theta) +\tilde \theta^2 \left(-2im^\Lambda+\frac14 \bar D^2\bar S^\Lambda\right).
\end{equation}
with $S^\Lambda$ as in \eqref{AltN_Xsol} and $W_{\alpha}^\Lambda = -\frac14 \bar D^2 D_\alpha U^\Lambda$, where $U^\Lambda$ is defined in \eqref{AltN_Vsol}.


\providecommand{\href}[2]{#2}\begingroup\raggedright\endgroup

\end{document}